\documentclass[twocolumn,showpacs,amsmath]{revtex4}
\usepackage{here}
\usepackage{bm}
\usepackage{epsfig,amsfonts,amsbsy}
\def\nn{\nonumber}
\def\lb{\label}
\def\ci{\cite}
\newcommand{\Ref}[1]{(\ref{#1})}
\def\a{\alpha}
\def\b{\beta}
\def\g{\gamma}

\def\e{\epsilon}
\def\o{\omega}
\def\th{\theta}
\def\s{\sigma}
\def\lam{\lambda}

\newcommand{\rd}{\mathrm{d}}
\newcommand{\p}{\partial}
\newcommand{\N}{\nabla}

\def\bra{\langle}
\def\ket{\rangle}
\def\l{\left}
\def\r{\right}

\def\f{\frac}
\def\cO{\mathcal O}

\def\bk{\bm k}
\def\bl{\bm l}

\begin{document}
\title{Interior of Black Holes and Information Recovery}
\author{Hikaru~Kawai\footnote{hkawai@gauge.scphys.kyoto-u.ac.jp}}
\affiliation {
Department of Physics, Kyoto University, Kyoto 606-8502, Japan}
\author{Yuki~Yokokura\footnote{yuki.yokokura@icts.res.in}}
\affiliation {International Centre for Theoretical Sciences, 
Survey No.151, Shivakote, Hesaraghatta Hobli, 
Bengaluru North - 560 089, India.}

\begin{abstract}
We analyze time evolution of a spherically symmetric collapsing matter from a point of view 
that black holes evaporate by nature.
We first consider a spherical thin shell that falls in the metric of an evaporating Schwarzschild black hole 
of which the radius $a(t)$ decreases in time. 
The important point is that 
the shell can never reach $a(t)$ 
but it approaches $a(t)-a(t)\f{\rd a(t)}{\rd t}$. 
This situation holds at any radius 
because the motion of a shell in a spherically symmetric system is not affected by the outside.  
In this way, we find that the collapsing matter evaporates without forming a horizon. 
Nevertheless, a Hawking-like radiation is created in the metric, 
and the object looks the same as a conventional black hole from the outside. 
We then discuss how the information of the matter is recovered.  
We also consider a black hole that is adiabatically grown in the heat bath 
and obtain the interior metric. 
We show that it is the self-consistent solution of 
$G_{\mu\nu}=8\pi G \langle T_{\mu\nu} \rangle$ 
and that the four-dimensional Weyl anomaly induces 
the radiation and a strong angular pressure. 
Finally, we analyze the internal 
structures of the charged and the slowly rotating black holes. 
\end{abstract}

\maketitle
\section{Introduction}
The picture of the black hole has changed significantly 
since the discovery of the Schwarzschild solution. 
In the classical level black holes are characterized by the existence of the event horizon, 
and objects which have entered into them cannot come back forever. 
In the quantum level, however, 
black holes evaporate in the vacuum \ci{Hawking} and 
can be in equilibrium with the heat bath of the Hawking temperature \ci{G-H}. 
One of the problems of this picture is the information paradox \ci{Hawking2}, 
which is essentially the disagreement between the information flow and the energy flow. 
Suppose we consider a process in which a black hole is formed by the collapse of matter. 
In the conventional picture, 
the matter crosses the horizon holding its own information. 
On the other hand, the Hawking radiation is thermal because it is created in vacuum, 
and it cannot reflect the detailed information of the collapsing matter. 
Therefore, if we assume that the evaporation occurs after the horizon is formed, 
we are forced to conclude that, although all the energy is emitted to infinity, 
the fallen information does not come back. 
In this paper, we reconsider the time evolution of the collapsing matter from 
the point of view that black holes are objects that evaporate inherently. 
Then, we find that 
no horizon appears and the matter is distributed in the whole region inside the black hole. 
Furthermore, we discuss that 
the Hawking radiation comes out through the matter and can exchange information with it. 

In order to understand the absence of the horizon, 
we first consider a spherical shell falling in the evaporating Schwarzschild metric. 
The important point is that 
the radius $a(t)$ is decreasing as $\f{\rd a(t)}{\rd t}=-\f{Cl_p^2}{a(t)^2}$ 
due to the Hawking radiation. 
Here $a=2GM$ is the Schwarzschild radius, $l_p=\sqrt{\hbar G}$ is the Planck length, 
and $C$ is a proportionality constant of order $\cO(1)$. 
Then, the shell does not catch up with the horizon completely
but approaches to $r=a(t)+\f{Cl_p^2}{a(t)}$. 
This is because 
the shell approaches the horizon in the time scale $a$, 
but during that time 
the radius $a$ decreases by $\Delta a = \l|\f{\rd a}{\rd t}\r| a =\f{Cl_p^2}{a}$. 

We then consider a spherically symmetric collapsing matter 
with a continuous distribution, and regard it as a set of thin shells. 
Because of the spherical symmetry 
the time evolution of each shell is not affected by the shells outside it. 
Then, the above argument can be applied to each shell. 
Thus, there is no trapped region,  
and no horizon appears. 
Nevertheless, we can show that 
a radiation is created from each shell, 
and it takes almost the same form as the conventional Hawking radiation. 
Interestingly, a strong angular pressure is also 
induced, against which the shell collapses and loses the energy. 

From these discussions it turns out that 
the evaporating object has a clear surface at $r=a(t)+\f{Cl_p^2}{a(t)}$ 
and that its interior is filled with matter and radiation 
while from the outside it looks almost the same as the conventional black hole. 
As we will see, no trans-Planckian problems occur
if the theory has many fields, for example about 100.

This picture of black holes 
indicates possible mechanisms of the information recovery. 
Since the Hawking radiation is produced near each shell, 
the radiation and the collapsing matter can interact and exchange the detailed information. 
The time scale of this process can be estimated as $\sim a \log \f{a}{l_p}$. 
Based on this mechanism, 
we discuss 
the possibility that 
the radiation depends on the initial information of the collapsing mater.

We also discuss  
a black hole which is adiabatically grown in the heat bath. 
We obtain its interior metric 
and show that it is the self-consistent solution of the semiclassical Einstein equation 
$G_{\mu\nu}=8\pi G \bra T_{\mu\nu} \ket$. 
Then, it is understood that 
the the four-dimensional Weyl anomaly 
produces the radiation and the angular pressure. 
We can also investigate the interior structure of the charged and the slowly rotating black holes. 

This paper is organized as follows. 
In Sec. \ref{general}, 
we explain the new picture of black holes. 
In Sec. \ref{3}, we discuss how the information comes back in the process of evaporation, 
and consider possible mechanisms of information recovery. 
In Sec. \ref{2}, we consider the black hole that is adiabatically grown in the heat bath. 
In Secs. \ref{4} and \ref{5}, we study the charged and the slowly rotating black holes. 
We give supplementary discussions in the Appendixes.

\section{New picture of black holes}\lb{general}

\subsection{Conventional picture of black holes}\lb{setup}
We review here the conventional picture of black holes. 
Suppose we consider a Schwarzschild black hole in the vacuum with a large mass 
$M=\f{a}{2G}$ compared with the Planck mass $m_p \equiv \sqrt{\f{\hbar}{G}}$. 
The black hole has the Hawking temperature and the Bekenstein-Hawking entropy \ci{Hawking,G-H,Bekenstein}, 
\begin{equation}\lb{T_S}
T_H=\f{\hbar}{4\pi a},~~~S_{BH}=\f{A}{4l_p^2}~,
\end{equation}
where 
$A=4 \pi a^2$ is the area of the horizon. 
From the Stephan-Boltzmann law \ci{Landau_S}, 
the time evolution of $a(t)$ can be expressed as
\begin{equation}\lb{da}
\f{\rd a}{\rd t}=-\f{2\s(a)}{a^2}~.
\end{equation}
Here $\s(a)$ is $l_p^2$ multiplied by a constant of order 1. 
$\s(a)$ depends on the detail of the theory, such as the number of species of fields \ci{foot1}. 
We expect that 
$\s(a)$ varies with $a$ slowly compared with $l_p$; $\f{\rd \s}{\rd a}l_p \ll \s$. 
Although these results were originally obtained by assuming the existence of the horizon, 
as we will see later, collapsing matter radiates and has the same entropy
even if there is no horizon.

Next we consider the smallest unit of energy following Bekenstein's argument \ci{Bekenstein}.  
Suppose we inject a wave packet of a massless particle with energy $\e$ to a black hole with radius $a$. 
In order for the wave to enter into the black hole, 
its wavelength needs to be smaller than the size of the black hole: 
\begin{equation}\lb{lam_B0}
\lam \lesssim a.
\end{equation}
Therefore, its energy should satisfy
\begin{equation}\lb{e_B0}
\e=\hbar \o = \hbar \f{2\pi}{\lam}\gtrsim \f{\hbar}{a}. 
\end{equation}
Thus, the minimum energy that can be added to a black hole with radius $a$ is given by
\begin{equation}\lb{e_B}
\e\sim \f{\hbar}{a}~,
\end{equation}
and the corresponding wavelength is  
\begin{equation}\lb{lam_B}
\lam \sim a~.
\end{equation}
Although the above argument is valid for massless particles, we can show the same results also for massive particles (see Appendix \ref{Bek_rev}).

The minimum energy \Ref{e_B} corresponds to 1 bit of information,
because the wavelength of the particle \Ref{lam_B} is as large as the black hole, and we have two possibilities, whether it goes inside the hole or not. 
Now, suppose we build up a black hole with radius $a$ from particles which have the minimum energy. 
Then, the total amount of the lost information is evaluated as 
\begin{align}\lb{S_Bek_est}
 S&\sim {\rm (the~number~of~processes)}\times {\rm (entropy~per~a~process)} \nonumber \\
 &\sim \f{a}{G\e}\times \log 2 \sim \f{a^2}{l_p^2}~,
\end{align}
which agrees with $S_{BH}$ in \Ref{T_S}. 
Note that if we use particles with $\e\gg \f{\hbar}{a}$ 
the entropy is much less than \Ref{S_Bek_est} (see Appendix \ref{Bek_rev}). 
As we will see later, 
the existence of the horizon is not essential for this estimation.

\subsection{Motion of a test particle near the evaporating black hole}
\begin{figure}[h]
 \begin{center}
 \includegraphics*[scale=0.2]{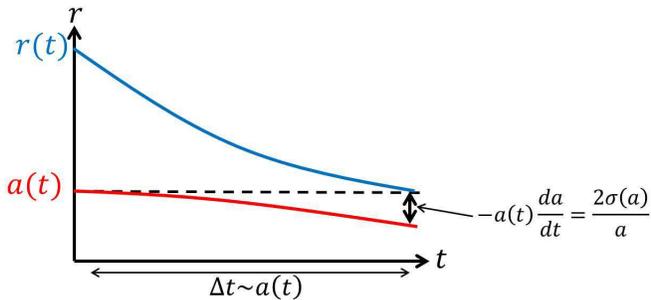}
 \caption{A test particle in the time-dependent Schwarzschild metric. 
 $r(t)$ cannot catch up with $a(t)$ as long as $\f{\rd a}{\rd t}(t)<0$. }
 \label{fig:surface}
 \end{center}
 \end{figure}

\lb{surface}
We analyze here the motion of a test particle near the evaporating Schwarzschild black hole. 
The outside spacetime can be approximately described by 
\begin{equation}\lb{Sch}
\rd s^2 = - \f{r-a(t)}{r}\rd t^2 + \f{r}{r-a(t)}\rd r^2 + r^2 \rd \Omega^2~, 
\end{equation}
where $a(t)$ satisfies \Ref{da} \ci{foot2}. 
If the test particle comes sufficiently close to $a(t)$, 
its radial coordinate $r(t)$ is determined irrespectively of its mass or angular momentum by
\begin{equation}\lb{r_t}
\f{\rd r(t)}{\rd t}=-\f{r(t)-a(t)}{r(t)}~.
\end{equation}
This is because any particle becomes ultrarelativistic near $r\sim a$ 
and behaves like a massless particle \ci{Landau_C} \ci{foot3}. 
From \Ref{r_t} we see that the particle approaches the radius $a$ in the time scale of $\cO(a)$. 
During this time, however, the radius $a(t)$ itself is slowly shrinking 
due to the Hawking radiation. 
Hence, the particle cannot catch up with the radius $a$ completely. 
Instead, $r(t)$ is always apart from $a(t)$ by $-a\f{\rd a}{\rd t}$. See Fig.\ref{fig:surface}. 

We can see this behavior explicitly by solving \Ref{r_t} as follows. 
Putting $r(t)=a(t)+\Delta r(t)$ in \Ref{r_t} and assuming $\Delta r(t)\ll a(t)$, 
we have 
\begin{equation}\lb{r_t2}
\f{\rd \Delta r(t)}{\rd t} = -\f{\Delta r(t)}{a(t)}-\f{\rd a(t)}{\rd t}~.
\end{equation}
The general solution of this equation is given by 
\begin{equation}
\Delta r(t)=C_0 e^{-\int^t_{t_0}\rd t' \f{1}{a(t')}} + \int^t_{t_0}\rd t' \l(-\f{\rd a}{\rd t}(t') \r)e^{-\int^{t}_{t'}\rd t''\f{1}{a(t'')}}, \nn
\end{equation}
where $C_0$ is an integration constant. 
Because $a(t)$ and $\f{\rd a}{\rd t}(t)$ can be considered to be constant in the time scale of ${\cal O}(a)$,
the second term can be evaluated as 
\begin{align*}
 &\int^t_{t_0}\rd t' \l(-\f{\rd a}{\rd t}(t') \r)e^{-\int^{t}_{t'}\rd t''\f{1}{a(t'')}}\\
 &\approx -\f{\rd a}{\rd t}(t) \int^t_{t_0}\rd t' e^{-\f{t-t'}{a(t)}}=-\f{\rd a}{\rd t}(t) a(t)(1-e^{-\f{t-t_0}{a(t)}}). 
\end{align*}
Therefore we obtain 
\begin{equation}
\Delta r(t)\approx C_0 e^{-\f{t-t_0}{a(t)}} -\f{\rd a}{\rd t}(t)a(t) (1-e^{-\f{t-t_0}{a(t)}}), \nn 
\end{equation}
which leads to 
\begin{align}\lb{R0}
r(t)&\approx a-a\f{\rd a}{\rd t}+Ce^{-\f{t}{a}} \nn \\
 &= a+\f{2\s(a)}{a}+Ce^{-\f{t}{a}}\longrightarrow a+\f{2\s}{a}~,
\end{align}
where $C$ is a positive constant and we have used \Ref{da} to obtain the second line. 
This result indicates that in the time scale of ${\cal O}(a)$ any particle approaches 
\begin{equation}\lb{R}
R(a)\equiv a +\f{2\s(a)}{a}~,
\end{equation}
and it will never cross the radius $a(t)$ as long as $a(t)$ keeps decreasing.
In the following we call $R(a)$ the surface of the black hole. 
We give a numerical demonstration of \Ref{R0} in Appendix \ref{R_numerical}. 

One might wonder if such a small radial difference $\f{2\s}{a}$ makes sense, 
since it looks much smaller than $l_p$. 
However, the proper distance between the surface and the horizon 
is estimated as 
\begin{equation}\lb{dl}
\Delta l = \sqrt{g_{rr}(R(a))}\f{2\s(a)}{a}\approx\sqrt{2\s(a)}~,
\end{equation}
because $g_{rr}(r)=\f{r}{r-a}$. 
In general this is proportional to $l_p$, but the coefficient can be large 
if we consider a theory with many species of fields.
In fact, in that case we have 
\begin{equation}\lb{largeN_0}
\s\sim N l_p^2\gg l_p^2~,
\end{equation}
where $N$ is the number of fields. 
We assume that $N$ is large but not infinite, for example, of the order of
100 as in the standard model.

So far, we have found the surface \Ref{R} based on the classical motion of particles.
However, we can show that the result is valid even if we treat the particles quantum mechanically.
Suppose that we throw a wave packet of a massless particle with frequency $\o$ to the black hole \ci{foot4}. 
Here, $\o$ is measured at $r\gg a$ and there the wavelength is given by $\lambda=\f{2\pi}{\o}$. 
Then, it becomes blueshifted as it approaches to $r=a+\Delta r$: 
\begin{equation}
\o_{local}=\f{\o}{\sqrt{-g_{tt}(r)}}~.\nn
\end{equation}
Thus, the local wavelength is given by 
\begin{equation}
\lambda_{local}=\f{2\pi}{\o_{local}}=\sqrt{-g_{tt}(r)}\lambda \approx \sqrt{\f{\Delta r}{a}}\lambda~.  \nn
\end{equation}
On the other hand, the proper distance $l$ from $r=a$ is given by
\begin{align}\lb{l}
l&=\int^{\Delta r + a}_a \rd r \sqrt{g_{rr}}= \int^{\Delta r + a}_a \rd r \f{1}{\sqrt{1-\f{a}{r}}} \nn\\
 &\approx \int^{\Delta r + a}_a \rd r \sqrt{\f{a}{r-a}}= 2\sqrt{a \Delta r}~. 
\end{align}
Therefore, in order that the wave is contained in a region with the size $l$, 
the wavelength needs to satisfy
\begin{equation}\lb{eik_con}
1 \lesssim \f{l}{\lambda_{local}}=\f{2a}{\lambda}~,
\end{equation}
which turns out to be the same as \Ref{lam_B0}. 
This tells that  
such a wave packet that can go into the black hole behaves as a particle near $r=a$. 
Therefore, we can conclude that \Ref{R} is the position where any wave approaches. 
Note that \Ref{eik_con} is nothing but the condition for the eikonal approximation. 

\subsection{New picture}\lb{story} 
\subsubsection{Surface of black holes}\lb{surface2} 
We consider spherically symmetric 
collapsing matter 
that forms a black hole in the conventional picture. 
As is discussed in Appendix \ref{many_model}, 
we can regard it as consisting of thin shells with the minimum energy \Ref{e_B}. 
An important point is that because of the spherical symmetry, 
the time evolution of each shell is not affected by the shells outside it, 
if we describe the motion using the local time. 
Then, assuming that the mass inside each shell is decreasing as in \Ref{da}, 
we can apply \Ref{R} for each shell, and find that it is always apart from its Schwarzschild radius.
Therefore, we can conclude that there is no trapped region. 

Next, we consider the outermost shell. 
Because its radius approaches $R(a(t))=a(t)+\f{2\s(a(t))}{a(t)}$,
all matter is stuffed in the region  $r<R(a(t))$, 
which means that the object has a clear boundary at $r=R(a(t))$. 
This is the reason why we call $R(a)$ the surface \ci{foot5}.
The region outside the surface is almost empty, 
and the object looks exactly the same as the conventional black holes 
when it is observed from the outside.
On the other hand, the inside of the surface is totally different.
In particular, the horizon no longer exists \ci{KMY,KY,Ho} 
because of the nontrivial distance \Ref{dl}. 
In the following, we call the object a black hole, although it is significantly different from the conventional ones.

Here, one might wonder if the Hawking radiation is really created by such an object. 
However, we can show that indeed it is \ci{KMY,KY} (see also Appendix \ref{many_model}). 
Generally, particle creation occurs in a time-dependent potential, and it takes the Planck-like distribution 
if the affine parameters on the null generators of the past and future null infinity are related exponentially \cite{Barcelo}. 
Indeed, in Appendix \ref{Der_T_H}, 
using the self-consistent metric obtained in Sec. \ref{2}, 
we show that particles are created from the vacuum in accordance with the Planck-like distribution with 
the Hawking temperature \ci{foot5_5} 
\begin{equation}\lb{T_H}
T_H(t)=\f{\hbar}{4\pi a (t)}~.
\end{equation}
However, as we will see in Sec. \ref{2_6}, 
the distribution may be modified 
by the interaction between the collapsing matter and the radiation. 

\subsubsection{Self-consistent time evolution of each shell}\lb{time_shell}
In this subsection we investigate how each shell loses energy during the Hawking radiation. 
When we analyze the time evolution of a shell, we can ignore the matter outside it because of the spherical symmetry, and 
regard the system as consisting of the shell and the core. 
Here, the core is the part of the system inside the shell, 
and we denote its radius and mass by $r'$ and $\f{a'}{2G}$, respectively. 
For simplicity, we assume that $r'$ is already very close to $R(a')$ and that
the Hawking radiation is emitted as a conventional black hole, 
\begin{equation}\lb{da'}
\f{\rd a'}{\rd t'}=-\f{2\s(a')}{a'^2}~,
\end{equation}
where $t'$ is the time without taking the matter outside the core into account. 
We also assume for simplicity that the shell has no thickness and denote its radius 
by $r_s$.  

\begin{figure}[t]
 \includegraphics*[scale=0.15]{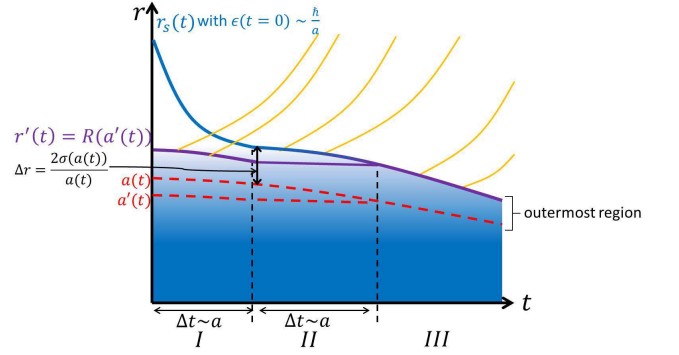}
 \caption{Time evolution of the core with $a'$ 
 and the shell with $\e(t=0)\sim\f{\hbar}{a}$ in the view of the local time $t$.}
 \label{fig:general}
 \end{figure}

Now we can discuss how the energy of the shell decreases from
its initial value, which we assume to be $\e\sim \f{\hbar}{a}$.
We can consider the following three stages. See Fig.\ref{fig:general}. 

\textit{Stage I}.---The shell is far from the core, 
and the radiation emitted from the core is not altered by the shell. 
The total mass $\f{a(t)}{2G}$ decreases as 
\begin{equation}\lb{da_I}
\f{\rd a}{\rd t}\approx-\f{2\s(a')}{a'^2}~.
\end{equation}

\textit{Stage II}.---In the time scale of order $a$,  
the shell comes close to the core. Then
the total system behaves like a black hole with radius $a$, 
the surface of which is located at $r=r_s=R(a)$, and radiates as usual \Ref{da}. 
However, the radiation comes mainly from the shell 
because the radiation from the core is extremely weakened by the redshift $\f{\rd t}{\rd t'}$ caused by the shell, 
although the core itself radiates constantly as \Ref{da'} in terms of $t'$. 
Therefore, $a'(t)$, and thus $R(a'(t))$, change very little during this stage. 

We can examine the time evolution of the energy of the shell more precisely, which is given by $\e(t)=\f{\Delta a(t)}{2G}$, 
where $\Delta a = a - a'$. 
The metric outside the core is given by 
\begin{equation}\lb{Sch2}
\rd s^2=\begin{cases}
 - \f{r-a(t)}{r}\rd t^2 + \f{r}{r-a(t)}\rd r^2+ r^2 \rd \Omega^2,~~~~~{\rm for}~r_s\leq r,\\
 - \f{r-a'(t')}{r}\rd t'^2 + \f{r}{r-a'(t')}\rd r^2 + r^2 \rd \Omega^2,~{\rm for}~R(a')\leq r\leq r_s,
\end{cases}
\end{equation}
where $a(t)$ and $a'(t')$ follow \Ref{da} and \Ref{da'}, respectively. 
The relation between $t$ and $t'$ is obtained as follows. 
First, we write the time evolution of $r_s$ in two ways using 
the metric outside and inside the shell: 
\begin{equation}\lb{r_eom_2}
\f{\rd r_s}{\rd t}= - \f{r_s(t)-a(t)}{r_s(t)},~~~\f{\rd r_s}{\rd t'}= - \f{r_s(t')-a'(t')}{r_s(t')}~.
\end{equation}
Then, by taking the ratio, we get 
\begin{align}\lb{tt'}
 \f{\rd t'}{\rd t} &= \f{r_s-a}{r_s -a'}=\f{\f{2\s(a)}{a}}{\f{2\s(a)}{a}+\Delta a}\\
 \lb{tt'2}
 &\approx 1-\f{a\Delta a}{2\s(a)}
\end{align}
where we have used $r_s=R(a)=a + \f{2\s(a)}{a}$ and assumed $\Delta a \lesssim \f{l_p^2}{a}$ together with \Ref{largeN_0}. 
By using \Ref{da}, \Ref{da'}, \Ref{tt'2}, and $\f{1}{a'^2}\approx\f{1}{a^2}\l(1+\f{2\Delta a}{a}\r)$, 
we obtain
\begin{align}\lb{e_decay_eq}
\f{\rd \e}{\rd t} &=\f{1}{2G}\l(\f{\rd a}{\rd t} -\f{\rd a'}{\rd t}\r)=\f{1}{2G}\l(\f{\rd a}{\rd t} -\f{\rd t'}{\rd t}\f{\rd a'}{\rd t'}\r) \nn\\
 &\approx -\f{1}{2G}\l[\f{2\s(a)}{a^2}-\l(1-\f{a\Delta a}{2\s(a)}\r)\f{2\s(a)}{a^2}\l(1+\f{2\Delta a}{a}\r) \r]\nn\\
 &\approx -\f{\e}{a}~,
\end{align}
which gives 
\begin{equation}\lb{e_decay}
\e(t)=\e(0)e^{-\f{t}{a(t)}}~.
\end{equation}
Here we have used the fact that 
$a(t)$ does not change significantly in the time scale of $\cO(a)$. 
Thus, the energy of the shell decreases exponentially in the time scale 
\begin{equation}\lb{t_decay}
\Delta t_{decay}\sim a~.
\end{equation}
Note that the redshift \Ref{tt'2} plays a crucial role in \Ref{e_decay_eq}.
This indicates that in general the radiation observed from the outside comes from the region near the surface.

\textit{Stage III}.---When the energy of the shell is exhausted, 
the core starts to radiate without redshift. 

Because the above argument can be applied to any shell, 
we conclude that the whole object evaporates from the outside as if an onion is peeled. 
A more detailed analysis is as follows.
First we estimate how many shells around the surface are moving without large redshift. 
From \Ref{tt'}, we have 
\begin{equation}\lb{time_flow}
\f{\rd t'}{\rd t} = \cO(1)\lesssim1 \Longleftrightarrow \Delta a \lesssim \f{\s}{a}~.
\end{equation}
Therefore, if we consider a black-hole-like object with radius $a_0$, 
the outermost region with the width 
\begin{equation}\lb{dr_surface}
\Delta r_{surface} \sim \f{\s(a_0)}{a_0}
\end{equation}
is not frozen, which contains $\f{\s}{l_p^2}$ shells with energy $\e\sim \f{\hbar}{a_0}$. 
Then, the lifetime of the object is estimated as 
\begin{align}\lb{a_consist}
 \Delta t_{life} &\sim ({\rm the~decay~time~of~a~shell})\nn\\
 &~~\times \f{({\rm the~total~number~of~the~shells}) }{({\rm the~number~of~shells~moving~at~the~same~time})}\nn\\
 &\sim a_0\times \l( \f{\f{a_0}{G}}{\e} \times \f{1}{\f{\s}{l_p^2}} \r) \sim \f{a_0^3}{\s}~,
\end{align}
which agrees with the one obtained from \Ref{da}.

\subsubsection{Interior of the evaporating black hole}
\begin{figure}[h]
 \includegraphics*[scale=0.165]{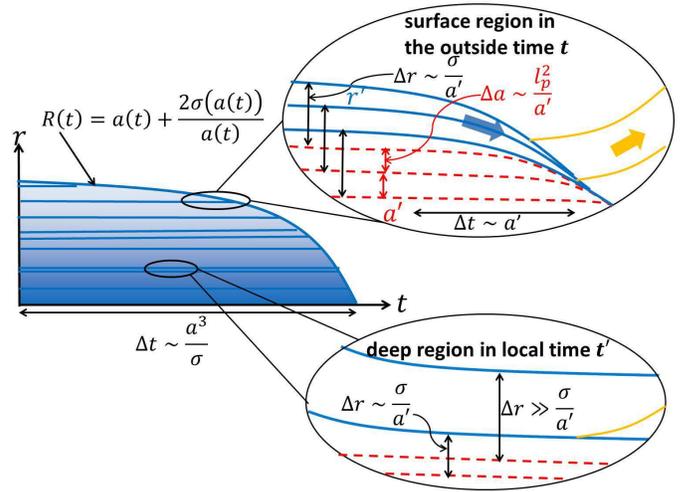}
 \caption{The interior picture of an evaporating black hole.}
 \label{fig:general2}
 \end{figure}

\lb{interior_eva}
Now we consider the interior of the object. 
See Fig.\ref{fig:general2}.
We examine the region deeper than the outermost one considered in the 
previous subsection. 
In terms of the local time, the shells in this region 
simply keep falling as \Ref{r_t} for the local quantities $r'$, $a'$, and $t'$ 
(see the lower closeup of Fig.\ref{fig:general2}). 
However, if we see them from the outside, 
time is frozen due to the large redshift 
after the outermost shells come close to the surface. 
To check this explicitly, 
we consider a shell with radius $r'$ in the deep region so that 
$\Delta a = a-a' \gg \Delta r_{surface}\sim \f{\s}{a}$. 
Here, $a$ and $a'$ are the Schwarzschild radius corresponding to the mass of the total system and 
that inside the shell, respectively. 
An important point here is that the shell has not necessarily reached $R(a')$. 
In fact, in that case, Eq. \Ref{tt'} becomes 
\begin{equation}\lb{tt'3}
\f{\rd t'}{\rd t}\approx \f{2\s}{a\Delta a}~,
\end{equation}
which leads to 
\begin{equation}\lb{freeze}
\Delta t' \sim a \Longleftrightarrow \Delta t \sim \f{a^2}{\s}\Delta a \gg a~.
\end{equation}
As we have seen, $\Delta t' \sim a $ is the time scale in which the shell reaches $R(a')$. 
On the other hand, 
$\Delta t \sim \f{a^2}{\s}\Delta a$ is the time scale 
in which the matter outside the shell evaporates by the Hawking radiation. 
Thus, we have seen that the interior region is almost frozen and 
its structure depends on the initial distribution \ci{foot6}.  
Each shell starts to evolve after the matter outside it disappears. 

Next, we discuss the outermost region with the width $\sim \f{\s}{a}$, 
where time flows without large redshift. 
For any initial distribution, each shell in this region 
reaches the asymptotic position $R(a')$ in the time scale $a$. 
Then, the Hawking radiation starts to be created, 
and the energy of the shell decreases exponentially 
as \Ref{e_decay} (see the upper closeup of Fig.\ref{fig:general2}). 
As we will see later, 
this time evolution depends on the initial data, 
which gives a natural mechanism of the information recovery.

\begin{figure}[h]
 \begin{center}
 \includegraphics*[scale=0.18]{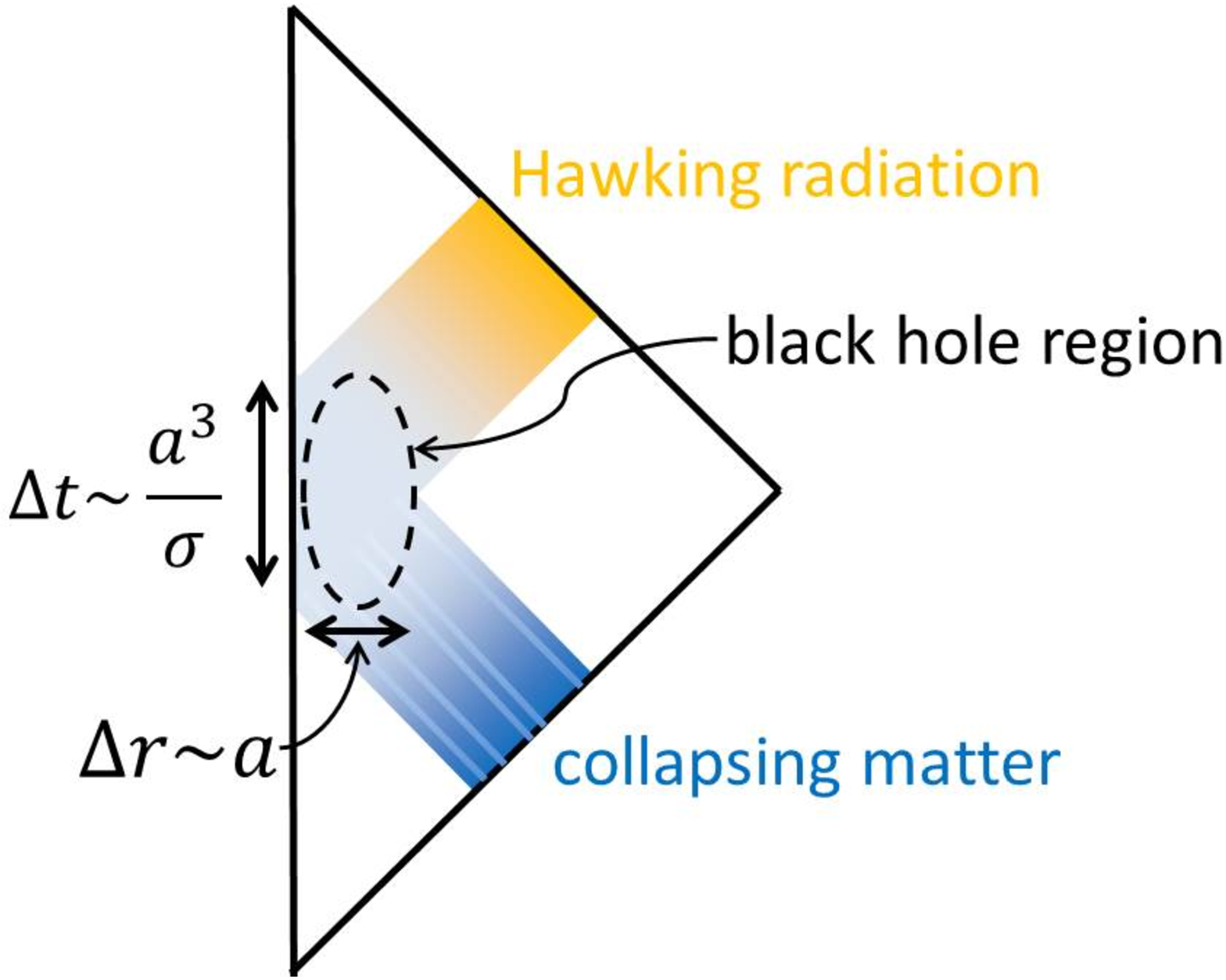}
 \caption{The Penrose diagram of the evaporating black hole in the vacuum. }
 \label{fig:Penrose_eva}
 \end{center}
 \end{figure}

So far, we have found that a collapsing matter 
becomes a compact object with the surface located at $r=R(a(t))$, 
and it evaporates without forming a trapped region in the time scale of $\f{a^3}{\s}$. 
We will also see in Sec. \ref{no_Planck} that no trans-Planckian problem occurs 
if the theory has many species of matter fields. 
Thus, we obtain the Penrose diagram as in Fig.\ref{fig:Penrose_eva}, 
which is topologically the same as the Minkowski space \ci{KMY, KY} \ci{foot7}. 

\subsection{Closer look at the surface and intensity $\s(a)$}\lb{2_6} 
In this subsection we examine the surface more precisely.
In particular, we consider the effect that some portion of the Hawking radiation is scattered 
back due to the gravitational potential or scattering with the other matters. 
\begin{figure}[h]
\includegraphics*[scale=0.13]{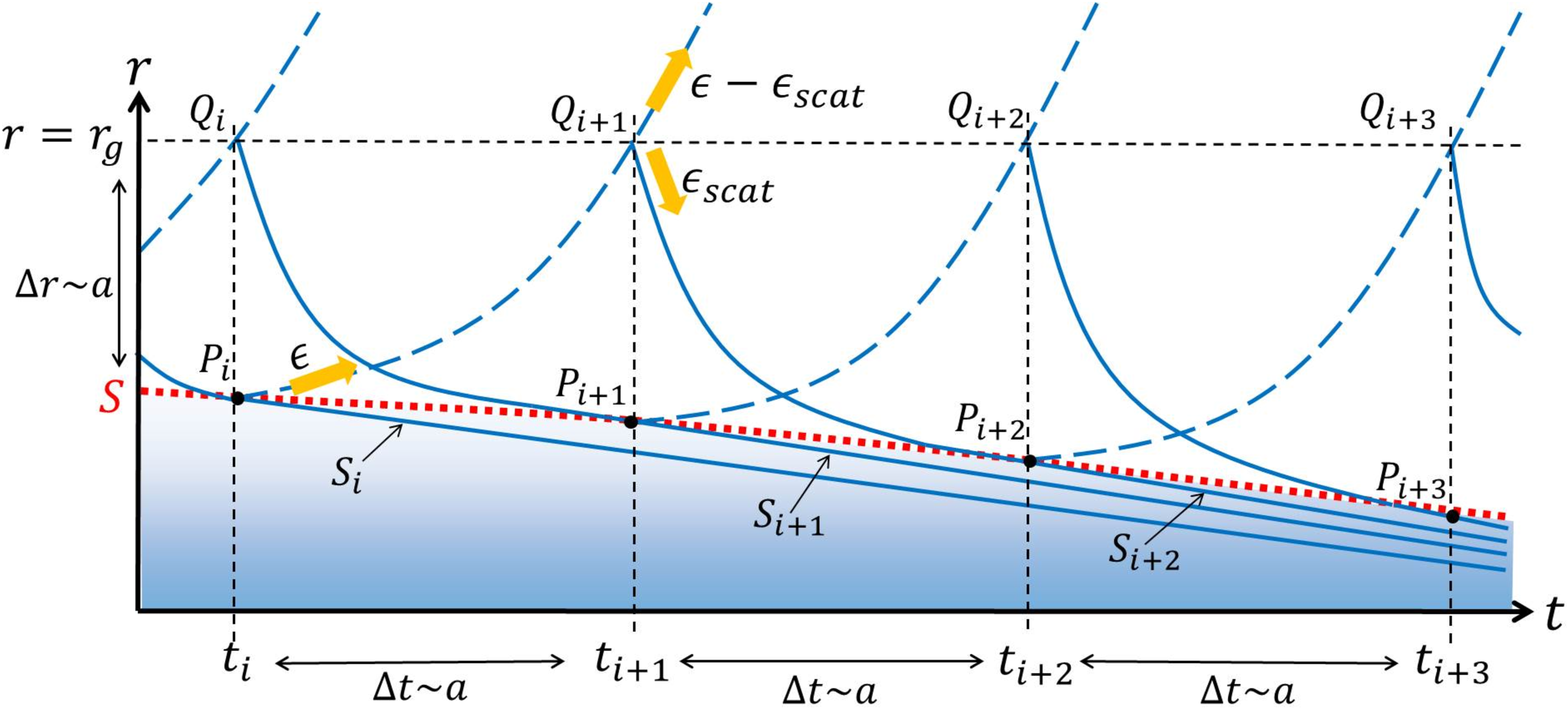}
\caption{A closer look at the outermost region. }
\label{fig:gray_new}
\end{figure}

The outermost region is magnified in Fig.\ref{fig:gray_new}. 
Here, for simplicity, time is discretized to the interval $\Delta t \sim a$. 
$S$ stands for the surface which is located at $r=R(a(t))=a(t)+\f{2\s(a(t))}{a(t)}$. 
Suppose that matter or radiation with energy $\e \sim \f{\s_0(a)}{a}$ is emitted from the surface at $t=t_i$, 
which is shown as $P_i$ in the figure \ci{foot8}. 
Here $\s_0(a)$ is the ``raw intensity" before taking the scattered flow 
into account.
In the time scale $a$, the matter reaches $r=r_g\equiv a+\Delta r$, 
where $\Delta r \sim a$. 
At this point $Q_{i+1}$, some amount of the matter is scattered back while the rest goes to infinity. 
In order to indicate the portion of the scattered energy, $\e_{scat}$, 
we introduce a function $g(a)$ as 
\begin{equation}\lb{g}
\e_{scat}=\f{g(a)}{1+g(a)}\e~.
\end{equation}
$g(a)$ should be an order 1 quantity 
and may depend on the matter Lagrangian. 
Then, in the time scale $a$, the scattered matter comes back to the surface $S$, 
which is shown as $P_{i+2}$. 
Then, it enters into the object and keeps going along the trajectory $S_{i+2}$, 
and it is covered by the subsequent matter or radiation. 
Here, $S_{i}$ is located at $r=R_0(a_i)\equiv a_i + \f{2\s_0(a_i)}{a_i}$, 
which is the asymptotic position of the shell that started from $Q_{i-1}$.
Thus, a part of the Hawking radiation comes back in the time scale $\sim a$. 
This process occurs continuously, 
and we find that the physical surface $S$ consists of the scattered matter and radiation. 

Now, we relate the raw intensity $\s_0$ to the net intensity $\s$. 
To do it, we estimate how the mass of the total system decreases during $t_{i+2}\leq t\leq t_{i+3}$. 
Adding the decrease by the raw emission and the increase by the scattered energy \Ref{g}, 
we have
\begin{align}\lb{g_est}
\Delta a &= -\f{2\s_0 }{a_{i+2}^2}\Delta t + \f{g}{1+g} \f{2\s_0}{a_{i}^2}\Delta t \nn \\
 &\approx  -\f{2\s_0 }{a^2}\Delta t + \f{g}{1+g} \f{2\s_0}{a^2}\Delta t = -\f{1}{1+g} \f{2\s_0}{a^2}\Delta t~. 
\end{align}
Comparing this with \Ref{da}, we obtain 
\begin{equation}\lb{g2}
\s(a)\equiv \f{\s_0(a)}{1+g(a)}~.
\end{equation} 
Although $g(a)$ may depend on the detail of the matter Lagrangian, 
the argument in Sec. \ref{surface} is still valid, 
and the position of the surface $S$ is determined by $\s$, as in \Ref{R}. 


\subsection{Stationary black holes in the heat bath}
\begin{figure}[h]
 \begin{center}
 \includegraphics*[scale=0.16]{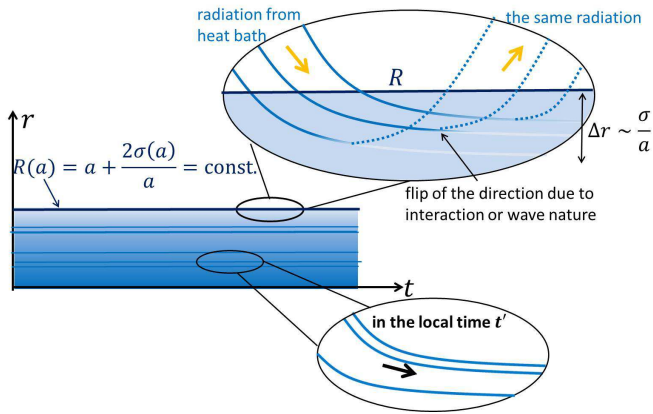}
 \caption{The black hole that is in equilibrium with the the heat bath. 
 The matter in the outermost region has been replaced with radiation from the heat bath.}
 \label{fig:surface_sta}
 \end{center}
 \end{figure}

\lb{BH_bath}
Here we consider how the black-hole-like object becomes in equilibrium 
with the heat bath. 
Suppose an evaporating object with mass $\frac{a}{2G}$ is put in the heat bath of temperature $T_H=\f{\hbar}{4\pi a}$. 
As the matter in the outermost region comes out of the object \ci{foot8}, 
the radiation from the heat bath replaces it. 
Because the collapsing matter is replaced by the radiation in the heat bath, 
this process is not an equilibrium one. 
After this process is completed in the outermost region, the system becomes 
stationary. 
Figure \ref{fig:surface_sta} represents this situation. 

\begin{figure}[h]
 \begin{center}
 \includegraphics*[scale=0.18]{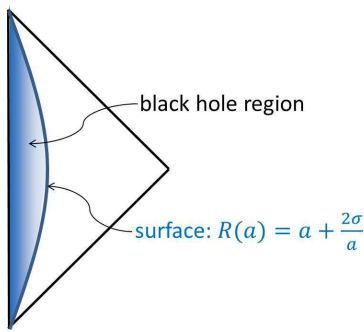}
 \caption{The Penrose diagram for the stationary black hole in Fig. \ref{fig:surface_sta}.}
 \label{fig:Penrose_sta}
 \end{center}
 \end{figure}

The remarkable point is that this process occurs only in the outermost region 
where time flows. 
The matter in the deeper region is almost frozen and keeps having its
initial information.
Nevertheless, the total object becomes in equilibrium with the heat bath 
in that the outgoing and ingoing flows balance almost completely. 
In the rigorous sense it is merely a stationary state, but in practice it 
behaves as an equilibrium state.
Finally, we show the Penrose diagram. See Fig. \ref{fig:Penrose_sta}.
While the interior structure depends on the initial distribution of matter,  
the exterior is universally described by the Schwarzschild metric, Eq. \Ref{Sch} with $a=$const..

\subsection{Absence of trans-Planckian problems}
\begin{figure}[h]
\begin{center}
\includegraphics*[scale=0.15]{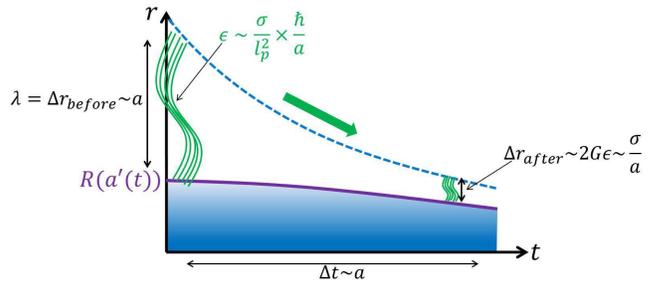}
\caption{ $\f{\s}{l_p^2}$ quanta with energy $\e_1\sim \f{\hbar}{a}$ approaching the core.}
\label{fig:no_Planck}
\end{center}
\end{figure}

\lb{no_Planck}
In this subsection we will show that the trans-Planckian effects are absent
if the theory has many fields. 
We consider the matter in the outermost region with width $\sim \f{\s}{a}$, 
where time flows as is discussed in \Ref{dr_surface}. 
The energy of the matter is given by $\e\sim \f{\s}{Ga}$ due to the relation $a=2GM$. 
Because the energy of the minimum quantum is $\e_1\sim\f{\hbar}{a}$ as discussed by Bekenstein \ci{Bekenstein},
we can regard the matter as consisting of 
$\f{\e}{\e_1}\sim \f{\s}{l_p^2}\sim N$ quanta. 
This is consistent with the fact that 
we have $N$ species of fields.  
In fact, we can 
consider 
how the $N$ quanta with wavelength $\sim a$ are compressed to form the outermost region just outside the core. 
See Fig. \ref{fig:no_Planck}. 
During this process, the waves are adiabatically compressed 
because the initial wavelength is $\sim a$, 
and the condition \Ref{eik_con} is satisfied. 

Each quantum with energy $\e_1\sim \f{\hbar}{a}$ is blueshifted as it falls toward the core,
and the energy becomes
\begin{equation}\lb{e1_loc}
\e_1^{local}=\f{\e_1}{\sqrt{-g_{tt}(R(a))}}\sim \f{\hbar}{\sqrt{\s(a)}}~.
\end{equation}
Here we have used \Ref{R} and \Ref{Sch2} to obtain 
$g_{tt}(R(a))=-\f{R(a)-a}{R(a)}\approx-\f{2\s(a)}{a^2}$. 
From \Ref{largeN_0}, we have $\e_1^{local} \ll m_p$ when $N$ is large. 
This indicates that each quantum does not have trans-Planckian energy. 

It should be noted that 
the wavelength $\lam^{local}_1$ of each quantum is of the same order 
as the proper length $\Delta l$ of the outermost region: 
\begin{align}\lb{lam_local_1}
\lam^{local}_1&=\f{\hbar}{\e_1^{local}}\sim \sqrt{\s(a)} \\
\Delta l &\sim\sqrt{g_{rr}(R(a))} \f{\s(a)}{a} \sim \sqrt{\s(a)}~.
\end{align}
Here we have used \Ref{R} and \Ref{Sch2} to obtain $g_{rr}(R(a))=\f{R(a)}{R(a)-a}\approx\f{a^2}{2\s(a)}$. 
This gives another support to the picture that 
the waves are adiabatically compressed as in Fig. \ref{fig:no_Planck} \ci{foot9}. 

The above results suggest that 
the typical scale in which things change is $\sim \sqrt{\s}$. 
Therefore, the curvature can be estimated as 
\begin{equation}\lb{R_est}
{\cal R}\sim \f{1}{(\sqrt{\s})^2}\sim \f{1}{\s}~,
\end{equation}
which is smaller than $l_p^{-2}$ from \Ref{largeN_0}.  
We will justify this result in Sec. \ref{2} by examining the self-consistent metric. See Eq. \Ref{curve}.

We can show that the energy density $\rho=-\bra T^t{}_t\ket$ is much smaller 
than the Planck scale if the object is sufficiently large, $a\gg l_p~$.
First, we note that for the general spherically symmetric system 
the Arnowitt-Deser-Misner energy (ADM energy) inside radius $r$ is given by \ci{Landau_C} 
\begin{equation}\lb{M_formula}
M(r)=4\pi \int_0^r \rd r' r'^2 \rho (r')~. 
\end{equation}
In Fig. \ref{fig:no_Planck}, the total energy $\e$ of the quanta is conserved, and 
using \Ref{M_formula}, we have 
\begin{equation}
\e\sim 4\pi a^2 \Delta r_{before } \rho_{before} \sim 4\pi a^2 \Delta r_{after} \rho_{after}~.
\end{equation}
Here, $\Delta r_{before}$, $\rho_{before}$, $\Delta r_{after}$, and $\rho_{after}$ 
are the size of the region and the energy density before and after the process, respectively. 
Using $\e\sim \f{\s}{Ga}$, $\Delta r_{before}\sim a$, $\Delta r_{after}\sim\f{\s}{a}$, and $N\sim \f{\s}{l_p^2}$, 
we obtain 
\begin{align}\lb{rho_b}
\rho_{before}&\sim \f{N\hbar}{a^4}~,\\
\lb{rho_a}
\rho_{after}&\sim \f{1}{Ga^2}~.
\end{align}
Both are much smaller than the Planck scale if $N$ is not too large.
Equation \Ref{rho_a} will be checked by the self-consistent solution [see Eq. \Ref{p}]. 

It is expected that the energy flux density is of the same order as the energy density, 
because the matter is ultrarelativistic near $R(a)$ as in \Ref{e1_loc}. 
Indeed, we can check this as follows. 
The Hawking flux is given by $J(a)=\f{\s(a)}{Ga^2}$ from \Ref{da}, 
and considering the double blueshift factors, we have 
\begin{align}\lb{J_shell}
J_{local}(R(a))&= \l(\f{1}{\sqrt{-g_{tt}(R(a))}}\r)^2J(a) \nn\\
 &\approx \f{a^2}{2\s(a)}\f{\s(a)}{Ga^2}=\f{1}{2G}~.
 \end{align}
Therefore, the flux per unit area is estimated as \ci{Ho}
\begin{equation}\lb{J_shell_loc}
j_{local}(R(a)) =\f{1}{4\pi R(a)^2}J_{local}(R(a)) \approx \f{1}{8\pi G a^2}~,
\end{equation}
which is again very small compared with the Planck scale.
Note that if the horizon  existed $J_{local}$ would diverge at $r=a$.

\subsection{Strong angular pressure}\lb{EMT_fire}
We show that a strong pressure in the angular direction
appears in the  interior of the black hole. 
We consider stage II of Fig. \ref{fig:general}, 
and evaluate the surface energy-momentum tensor on the shell. 
We can use the junction condition for a null hypersurface \ci{Israel_null, Poisson}, 
because the shell moves almost lightlike along \Ref{r_t}. 
We obtain the surface energy density and surface pressure \ci{KMY} (see Appendix \ref{Israel_con} for the derivation.): 
\begin{equation}\lb{EMT_shell}
\e_{2d}=\f{\e}{4\pi r_s^2},~~
p_{2d}= \f{-r_s}{8\pi G (r_s-a)^2}\l[\f{\rd a}{\rd t}-\l(\f{r_s-a}{r_s-a'}\r)^2 \f{\rd a'}{\rd t'} \r]~.
\end{equation}
$\e_{2d}$ simply represents the energy per unit area of the shell with energy $\e$. 
In the expression of $p_{2d}$, 
the first term corresponds to the total energy flux from the whole object \Ref{da}.
The second term represents the energy flux from the core \Ref{da'} 
that is redshifted due to the shell [see Eq. \Ref{tt'}]. 
Thus, $p_{2d}$ is induced by the Hawking radiation from the shell itself. 

We can estimate $p_{2d}$ for $r_s=R(a)$  
using a similar argument to \Ref{e_decay_eq} (see Appendix \ref{Israel_con}): 
\begin{align}\lb{p_2d_shell}
p_{2d}&\approx \f{r_s}{8\pi G (r_s-a)^2}\l[\f{2\s }{a^2}-\l(1-\f{a\Delta a}{\s}\r) \f{2\s}{a^2}\l(1+2\f{\Delta a}{a}\r) \r]\nn\\ 
 &\sim \f{a}{G N^2 l_p^2}~.
\end{align}
Here $\Delta a \sim \f{l_p^2}{a}$ and \Ref{largeN_0} have been used. 
Let us see how large this is as a three-dimensional quantity.  
Noting that 
$p_{2d}$ is a force per unit length on the spherical shell with the curvature radius $\sim a$, 
we expect that the  three-dimensional pressure is given by 
\begin{equation}\lb{p_3d}
p_{\th}\sim \f{p_{2d}}{a}\sim \f{1}{GN^2l_p^2}~. 
\end{equation} 
As we will show in Sec. \ref{2_5}, 
Eq. \Ref{p_3d} can be understood by the four-dimensional Weyl anomaly \ci{foot11}. 
This is large, but not trans-Planckian because of \Ref{largeN_0}. 
If an observer falls into the object as $S_i$ in Fig. \ref{fig:gray_new},
he will find this intense pressure around the surface. 
This may be identified with the firewall \ci{firewall}, although
the interpretation is rather different. 
Furthermore, the pressure is extremely anisotropic 
because the radial pressure $p_r$ can be estimated as 
$p_r \sim j_{local} \sim \f{1}{Ga^2}$ from \Ref{J_shell_loc}, 
which is much smaller than \Ref{p_3d}. 
Therefore the interior of the object cannot be regarded as the ordinary fluid.

At a first glance the strong angular pressure seems mysterious. 
However, it plays an important role to decelerate and sustain the collapsing matter. 
They lose energy as they shrink against the pressure, and the energy is converted
to the Hawking radiation. 
Therefore, we can conclude that the existence of the strong angular 
pressure is self-consistent and robust.
In this sense, the new picture is very different from that of the two-dimensional 
models \ci{RST}. 

\section{Information recovery in the new picture}\lb{3}

\subsection{Interaction}\lb{3_3}
As we have seen in the new picture, the Hawking radiation is created near the surface (see Appendix \ref{Der_T_H} for the detailed analysis).
Therefore, it is important to consider the interaction between the collapsing matter and the Hawking radiation. 
Here, we estimate the time scale of the scattering by
considering only the s-wave and approximating the interaction
as a one-dimensional scattering problem.

\begin{figure}[h]
 \includegraphics*[scale=0.17]{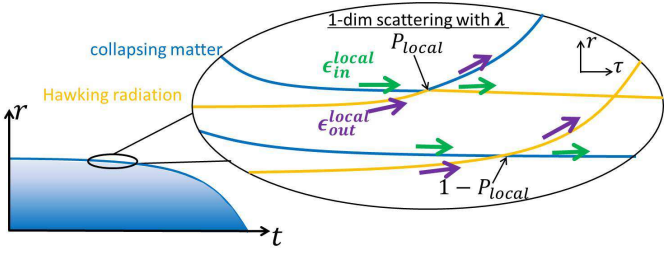}
 \caption{Scattering process between ingoing matter and outgoing radiation 
 in the outermost region.} 
 \label{fig:scattering}
\end{figure}

Suppose that the ingoing matter and outgoing Hawking radiation interact
with a small dimensionless coupling constant $\lambda$ (see Fig. \ref{fig:scattering}).
There are two possible cases.
One is the backward scattering, after which the ingoing matter goes outward
and the Hawking radiation inward. 
In this case, they exchange their energies.
The other is the forward scattering, 
in which the matter and radiation go through each other. 
If we denote the probability of the backward scattering per unit proper time by
$P_{local}$, the scattering proper time $\Delta \tau_{scat}$ is evaluated from 
\begin{equation}\lb{Def_scat}
\int_0^{\Delta \tau_{scat}} \rd \tau P_{local }=1 ~,
\end{equation}
where
$\rd \tau=\sqrt{-g_{tt}(r\approx R(a))}\rd t \approx \f{\sqrt{2\s(a)}}{a} \rd t$. 

We can estimate $P_{local}$ as follows. First we have
\begin{equation}\lb{Pro}
P_{local}=\Omega_{cross} \times F~,
\end{equation}
where $\Omega_{cross}$ and $F$ are the cross section and
the number flux, respectively.
$\Omega_{cross}$ is given by 
\begin{equation}\lb{cross0}
\Omega_{cross} \sim \lambda^2 \f{\hbar}{\e_{in}^{local}} \f{\hbar}{\e_{out}^{local}}~,
\end{equation}
where $\e_{in}^{local}$ and $\e_{out}^{local}$ are 
the local energy of the ingoing matter and outgoing radiation, respectively. 
We note that the energy of the wave in the collapsing matter decays as \Ref{e_decay} 
and that the typical local energy scale is given by \Ref{e1_loc}. 
Thus, we have 
\begin{equation}\lb{cross}
\Omega_{cross} \sim \lambda^2 \s(r) e^{\f{\tau}{\sqrt{2\s(r)}}}~.
\end{equation}
Next, using \Ref{J_shell_loc}, 
we have 
\begin{equation}\lb{F}
F \sim j_{local}(R(a)) \times \f{1}{\e_{out}^{local}} \sim \f{\sqrt{\s(a)}}{l_p^2 a^2}~.
\end{equation}
Then, Eq. \Ref{Def_scat} becomes 
\begin{align*}
1 &= \int_0^{\Delta \tau_{scat}} \rd \tau \Omega_{cross}\times F \\
 &\sim \int_0^{\Delta \tau_{scat}}  \rd \tau  \lambda^2 \s(a)  e^{ \f{\tau}{\sqrt{2\s(a)}} } \times \f{\sqrt{\s(a)}}{l_p^2a^2}\nn\\ 
&\approx \f{\sqrt{2}\lambda^2 \s(a)^2}{l_p^2a^2}e^{\f{\Delta \tau_{scat}}{\sqrt{2\s(a)}}}~,
\end{align*}
from which we obtain 
\begin{equation}\lb{tau_scat}
\Delta \tau_{scat}\sim \sqrt{2\s(a)} \log \l[\f{l_p a}{\lambda \s(a)} \r]~.
\end{equation}
In terms of the Schwarzschild time, it corresponds to 
\begin{equation}\lb{t_scat}
\Delta t_{scat}\sim a \log \l[\f{l_pa}{\lambda \s(a)} \r]~. 
\end{equation}

\subsection{Wave nature of matter}\lb{wave_nature}
We next examine another possible mechanism of the information recovery. 
Equation \Ref{e_decay} indicates that 
the wavelength of the particles in the outermost region increases as
\begin{equation}\lb{w_wave}
\lambda_{local}(\tau)=\f{\hbar}{\e_{local}(\tau)} \sim \sqrt{\s(a)}e^{\f{\tau}{\sqrt{2\s(a)}}}~,
\end{equation}
where we have used \Ref{e_decay}, \Ref{lam_local_1}, and  $\rd \tau\approx \f{\sqrt{2\s(a)}}{a} \rd t$. 
The matter can no longer stay in the black hole
if the wavelength becomes larger than the size of the black hole:
\begin{equation}\lb{QM_con}
\lambda_{local}(\tau) \gtrsim l_{BH}(\tau)~.
\end{equation}
Here,
\begin{equation}\lb{l_BH}
l_{BH}(t)=\int_0^{R(a(t))}\rd r \sqrt{g_{rr}(t,r)}
\end{equation}
is the proper size of the black hole which depends on the interior metric $g_{rr}(t,r)$. 
We can rewrite the condition \Ref{QM_con} by using \Ref{w_wave} as 
\begin{equation}\lb{tau_wave0}
\tau \gtrsim \sqrt{2\s(a)}\log \l[\f{l_{BH}}{\sqrt{\s(a)}}\r]\equiv \Delta \tau_{wave}~.
\end{equation}
Thus, $\Delta \tau_{wave}$ is the time scale in which 
the matter comes back by the wave nature.

$l_{BH}$ is estimated as follows.
We first note that 
the flat space, in which $g_{rr}=1$, has the minimum proper length. 
On the other hand, we expect that 
the interior of the adiabatically formed black hole, which we will see in the next section, 
has the longest proper length. 
Thus, using the self-consistent metric \Ref{sta_metric}, we have 
\begin{equation}
1 \leq g_{rr}(r) \leq \f{r^2}{2\s(r)}\Longrightarrow 
a \lesssim l_{BH}(a)\lesssim \f{a^2}{2\sqrt{2\s(a)}}~.
\end{equation}
At any rate, $\Delta \tau_{wave}$ is approximately given by
\begin{equation}\lb{tau_wave}
\Delta \tau_{wave}\sim \sqrt{2\s(a)}\log \f{a}{\sqrt{\s(a)}}~,
\end{equation}
from which we obtain the time scale in terms of the Schwarzschild time
\begin{equation}\lb{t_wave}
\Delta t_{wave}(a)\sim a\log \f{a}{\sqrt{\s(a)}}~. 
\end{equation}

\subsection{Time evolution of information recovery}\lb{info} 
Based on the above results, 
we can discuss how the information recovers in the evaporation process. 
Because the two time scales \Ref{t_scat} and \Ref{t_wave} are essentially the same
and expressed as
\begin{equation}\lb{t_back}
\Delta t_{back} \sim a \log \f{a}{l_p}~,
\end{equation}
we do not have to distinguish the detailed mechanism.
At any rate it indicates that after the particle with initial energy $\sim \f{\hbar}{a}$ reaches the surface of the black hole, 
it comes back with its initial information in the time scale $\Delta t_{back}~$. 
Thus the energy flow agrees with the information flow, 
and the information comes back in sequence from the outside as the black hole evaporates. 
Note that \Ref{t_back} also corresponds to the thermalization time 
in the sense of Sec. \ref{BH_bath}.

We can examine the time evolution of the entanglement entropy 
between the black hole and the emitted matter. 
Suppose that initially the collapsing matter is in a pure state. 
The time evolution of the total system is unitary 
because there is no trans-Planckian physics and it is described by a local field theory. 
Therefore, as usual, the entanglement entropy increases for a while, 
and it starts to decrease after about half of the black hole has evaporated. 
This time is about the half of the lifetime of the black hole $\sim \f{a^3}{\s}$, 
which is essentially the same as the Page time \ci{Page}. 
When the black hole evaporates completely, the entropy becomes zero again, 
which means that all the information has come out.

\subsection{Nonconservation of baryon number}\lb{baryon}
We discuss the conservation of the baryon number in the evaporation process. 
Suppose we construct a black hole with radius $a$ from $\f{a^2}{l_p^2}$ baryons 
by repeating Bekenstein's operation \ci{Bekenstein} (see Appendix \ref{Bek_rev}). 
The important point is that although each baryon has the rest mass $m$,
it increases the ADM mass of the black hole by $\sim\f{\hbar}{a}~$, which is much
smaller than $m$.
Then, if the baryon number is conserved, not all the baryons can come back to infinity after the evaporation
because the total rest mass $\sim \f{a^2}{l_p^2} m$ is much larger than 
the total ADM mass $\f{a}{2G}$. 
In other words, even if the baryons are emitted near the surface, 
they cannot reach infinity due to the binding energy. 
String theory may give an answer to this paradox.
Actually in string theory it is believed that there is no continuous global symmetry, 
and any continuous symmetry must be gauged \ci{string}. 
Therefore, we can expect that 
baryons are converted to massless particles through some interaction 
in the region near the surface where the local energy of particles is close to the 
Planck scale as \Ref{e1_loc}. 
Thus the black hole made from baryons can evaporate by 
violating the baryon number conservation.

\section{Adiabatically formed Schwarzschild black hole}\lb{2}
\subsection{Black hole in the heat bath}
\begin{figure}[h]
\includegraphics*[scale=0.168]{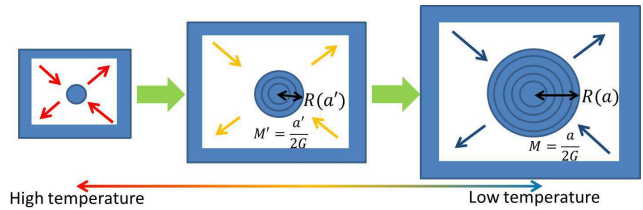}
\caption{The black hole that is formed adiabatically in the heat bath.}
\label{fig:adiaBH}
\end{figure}

\lb{2_1}
We consider a small Schwarzschild black hole put in a heat bath and grow it 
to a large one adiabatically 
by changing the temperature and size of the heat bath properly 
(see Fig. \ref{fig:adiaBH}). 
As was discussed in Sec. \ref{general}, 
the black hole has the surface at $r=R(a')$
when the mass is $\f{a'}{2G}$, 
and the radiation from the black hole always balances that from the heat bath as in Fig. \ref{fig:surface_sta}. 
Therefore the structure at the radial coordinate $r$ is completely determined
when the surface is at $r$. Once it is determined, it is not altered
even after the black hole grows
because of the spherical symmetry \ci{foot12}. 
Hence, the interior metric is independent of the total size of the black hole, $a$. 
Thus, we can parametrize the metric as
\begin{equation}\lb{ansatz}
\rd s^2 = - \f{1}{B(r)}e^{A(r)}\rd t^2+ B(r) \rd r^2 + r^2 \rd \Omega^2,~~~{\rm for}~~r\leq R(a)~,
\end{equation}
where neither $A(r)$ nor $B(r)$ depends on $a$. 
In the next subsection, we will determine $A(r)$ and $B(r)$ in terms of two phenomenological functions.
On the other hand, the region $r\geq R(a)$ is approximated by the Schwarzschild
metric, Eq. \Ref{Sch} with $a=$const..

We discuss here the time scale for adiabaticity. 
As we have seen in the previous sections, the unfrozen region at each stage has the width $\sim \f{\s(a')}{a'}$, \Ref{dr_surface},
and it is thermalized 
in the time scale $a'\log \f{a'}{l_p}$, \Ref{t_back}.
Therefore, if the radius $a$ changes by $\Delta a \sim \f{\s}{a}$ in 
a time scale longer than $a\log \f{a}{l_p}$, 
it can be regarded as an adiabatic process \ci{foot13}.

\subsection{Determination of the interior metric}\lb{2_3}
The function $B(r)$ is easily determined
if we assume that the metric at radial coordinate $r$ is completely
frozen to the value when the surface is at $r$ in the growing process Fig. \ref{fig:adiaBH}.
First, $g_{rr}$ on the surface is obtained from \Ref{Sch} by setting $r=R(a)$:
\begin{equation}\lb{grr}
g_{rr}|_{r=R(a)}=\f{R(a)}{R(a)-a}=\f{R(a)a}{2\s(a)}\approx \f{R(a)^2}{2\s(R(a))}~. 
\end{equation}
In the last expression we have replaced $a$ with $R(a)$ 
because $\f{2\s}{a}$ is much less than $a$ for a large black hole, $a\gg l_p$. 
This can be directly identified with $B(R(a))$ in \Ref{ansatz} because the radial coordinate $r$ 
is uniquely fixed in the Schwarzschild coordinate \Ref{Sch}: 
$B(R(a))=\f{R(a)^2}{2\s(R(a))}~.$
Because this result holds for any $a$, and we have postulated that $A(r)$ and $B(r)$ do not depend on $a$,  
we find that the function $B(r)$ is determined as
\begin{equation}\lb{B}
B(r)=\f{r^2}{2\s(r)}~.
\end{equation}

In order to determine $A(r)$, 
we consider the energy-momentum flow inside the black hole. 
Because the system is in equilibrium, it has the time-reversal symmetry 
and satisfies
\begin{equation}\lb{T1}
-\bra T^{\mu \nu} \ket k_\nu=\eta (l^\mu+f(r) k^\mu),~~~-\bra T^{\mu \nu}\ket l_\nu=\eta (k^\mu+f(r) l^\mu)~. 
\end{equation}
Here, $f(r)$ is expected to be of order 1 and vary slowly compared with $l_p$:   
$\f{\rd f}{\rd r}l_p \ll f$. 
$\bl$ and $\bk$ are the radial outgoing and ingoing null vectors, respectively,
\begin{equation}\lb{l-k}
\bl = e^{-\f{A}{2}}\p_t+\f{1}{B}\p_r ,~~~\bk = e^{-\f{A}{2}}\p_t-\f{1}{B}\p_r ~,
\end{equation}
which transform under time reversal as $(\bl,\bk)\rightarrow (-\bk,-\bl)$. 
Equation (\ref{T1}) can be rewritten as   
\begin{equation}\lb{T2}
\bra T^{\bk \bk}\ket:\bra T^{\bl \bk}\ket=1:f,~~~\bra T^{\bk \bk}\ket=\bra T^{\bl \bl}\ket~,
\end{equation}
where $T^{\bk \bk}$ stands for $T^{\mu \nu}k_\mu k_\nu$, and so on. 
This is also expressed in terms of  
the ratio between the energy density $-\bra T^t{}_t \ket$ and the radial pressure $\bra T^r{}_r \ket$: 
\begin{equation}\lb{T3}
\f{\bra T^r{}_r \ket}{- \bra T^t{}_t \ket}=\f{1-f}{1+f}~.
\end{equation}

\begin{figure}[h]
\includegraphics*[scale=0.22]{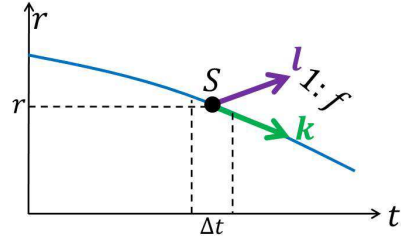}
\caption{The meaning of $f(r)$.}
\label{fig:f}
\end{figure}

Here we discuss the physical meaning of $f(r)$ (see Fig.\ref{fig:f}). 
The vector $P^\mu=\bra T^{\mu \bk}\ket$ at $r$ inside the black hole represents the energy-momentum flow 
through the ingoing lightlike spherical surface $S$ of radius $r$. 
Since $S$ can be regraded as an evaporating black hole with $M\approx \f{r}{2G}$, 
$P^\mu$ describes the radiation from the black hole. 
If the radiated particle is massless and propagates outward along the radial direction without scattering, 
$P^\mu$ should be parallel to $l^\mu$, which means $f=0$. 
Therefore, the value of $f$ represents the deviation from such an ideal situation. 
If the radiated particle is massive, $P^\mu$ is timelike, and we have $f>0$. 
Even when the particle is massless, $f$ can become nonzero 
if the particle is scattered in the ingoing direction by gravitational potential or 
interaction with other particles. 
Note that $f(r)$ in the outermost region may be different from one in the deeper region. 
This is because the energy density outside the surface 
is less than that in the deeper region [see \Ref{rho_b} and \Ref{rho_a}], 
and the probability of scattering should be different. 
In the following, we assume that $f(r)$ does not depend on $a$ except for 
a very thin outermost region.

Once $f(r)$ is given, we can determine $A(r)$ as follows.
Using \Ref{T3} and the Einstein equation we obtain
\begin{equation}\lb{dA}
\f{2}{1+f} =\f{G^r{}_r}{-G^t{}_t}+1=\f{r\p_r A}{B-1+r\p_r \log B}\approx \f{r\p_r A}{B}~. 
\end{equation}
In the last equation, we have used $B\gg 1$ and $B\gg r\p_r \log B$ for $r \gg l_p$, 
which can be easily checked from \Ref{B}. From \Ref{B} and \Ref{dA} we have
\begin{equation}\lb{A}
A(r)=\int^r_{r_0}\rd r'\f{r'}{(1+f(r'))\s(r')}~,
\end{equation}
where $r_0$ is a reference point.

Now, by connecting the inside metric \Ref{ansatz} and the outside metrics \Ref{Sch} at the surface, 
we can write down the metric of the black hole of radius $a$ that has been grown in the heat bath \ci{KY}, 
\begin{widetext}
\begin{equation}\lb{sta_metric}
\rd s^2=\begin{cases}
-\f{2\s(r)}{r^2} e^{- \int^{R(a)}_r \rd r' \f{r'}{(1+f(r'))\s(r')}} \rd t^2 + \f{r^2}{2\s(r)} \rd r^2 + r^2 \rd \Omega^2,~~{\rm for}~~r\leq R(a)~,\\
 - \f{r-a}{r}\rd t^2 + \f{r}{r-a}\rd r^2 + r^2 \rd \Omega^2,~~{\rm for}~~r\geq R(a)~,
\end{cases}
\end{equation}
\end{widetext}
where $R(a)=a + \f{2\s(a)}{a}~$.
This metric is continuous at $r=R(a)$. 
We emphasize that the interior metric of 
\Ref{sta_metric} does not exist in the classical limit $\hbar \rightarrow 0$ because $\s(r)$ vanishes.

As we have discussed in Sec. \ref{general}, in general the inside metric
depends on the initial distribution of the collapsing matter. If we put such an object
in the heat bath with the Hawking temperature, the outermost region becomes
in equilibrium (as discussed in Sec. \ref{BH_bath}) in the time scale $a \log\frac{a}{l_P}$, 
while the deeper region is almost frozen. 
However, if we wait a very long time, or if we make it shrink and grow
adiabatically by changing the temperature of the heat bath, 
the inside is replaced by the radiation from the heat bath, and the object becomes exactly in equilibrium, 
the metric of which is given by \Ref{sta_metric}. 
In this sense, the metric \Ref{sta_metric} represents the state with the maximum entropy. 
We will show later that it agrees with the area law. 

Similarly, we can construct the metric of the evaporating black hole in the vacuum. 
To do that, we first rewrite \Ref{ansatz} in the Eddington-Finkelstein-like coordinates as
\begin{equation}\lb{EF}
\rd s^2 = - e^{\f{A(r)}{2}}\l(\f{1}{B(r)}e^{\f{A(r)}{2}}\rd u+2\rd r \r)\rd u + r^2 \rd \Omega^2~,
\end{equation}
where we have introduced $u$-coordinate by $\rd u = \rd t - B(r)e^{-\f{A(r)}{2}}\rd r$. 
Then, we obtain the metric by connecting \Ref{EF} to the Vaidya metric \ci{Vaidya} along the null surface $S$, 
\begin{widetext}
\begin{equation}\lb{eva_metric}
\rd s^2=\begin{cases}
-e^{- \int^{R(a(u))}_r \rd r' \f{r'}{2(1+f(r'))\s(r')}} \l(\f{2\s(r)}{r^2} e^{- \int^{R(a(u))}_r \rd r' \f{r'}{2(1+f(r'))\s(r')}} \rd u 
+ 2\rd r \r) \rd u + r^2 \rd \Omega^2,~~{\rm for}~~r\leq R(a(u))~,\\
 - \f{r-a(u)}{r}\rd u^2 -2\rd r \rd u + r^2 \rd \Omega^2,~~{\rm for}~~r\geq R(a(u))~,
\end{cases}
\end{equation}
\end{widetext}
where $R(a(u))=a(u) + \f{2\s (a(u))}{a(u)}$, and $a(u)$ satisfies $\f{\rd a(u)}{\rd u}=-\f{\s(a(u))}{a(u)^2}$. 
This metric is continuous at $r=R(a(u))$.

\subsection{Consistency checks}\lb{2_4}
We give some consistency checks here.
First, we investigate the large redshift inside the black hole.
The $tt$-component of \Ref{sta_metric} behaves as  
$-g_{tt}\sim \exp \l(-\f{a}{(1+f)\s}(R-r)-2\log \f{a}{\sqrt{2\s}} \r)$ slightly below the surface, $r \lesssim R$. 
Here, we have used the fact that $\s(r)/l_p^2$ and $f(r)$ are of order 1 
and small compared with $r/l_p$. 
This means that time flows only in the outermost region 
with the width of ${\cal O}\l( \f{l_p^2}{a}\r)$,  
and it is exponentially frozen in the deeper region, 
which is consistent with \Ref{dr_surface}.  

Next, we examine the validity of the use of the Einstein equation. 
From \Ref{sta_metric}, we can evaluate the geometrical invariants in the region $l_p \lesssim r \leq R(a)$ and obtain 
\begin{equation}\lb{curve}
R,~\sqrt{R_{\mu\nu}R^{\mu\nu}},~\sqrt{R_{\mu\nu\a\b}R^{\mu\nu\a\b}}\sim \f{1}{(1+f)^2 \s}~.
\end{equation}
This  means that if the condition 
\begin{equation}\lb{large_N}
\s(1+f)^2\gg l_p^2~
\end{equation}
is satisfied, the curvature is small compared to $l_p^{-2}$, 
and we can use the Einstein equation without worrying about the higher-derivative corrections. 
Note that \Ref{large_N} is the same as \Ref{largeN_0} because $f=\cO(1)$, 
and \Ref{curve} is consistent with \Ref{R_est}. 
The condition \Ref{large_N} is realized if there are many fields as
in the standard model.
Although this field-theoretic approach does not apply to the small region $0\leq  r \lesssim l_p$, 
the curvature at $r\approx l_p$ is smaller than the Planck scale as \Ref{curve}, and 
dynamics in such a small region would be resolved by string theory. 
In this sense, this metric does not have a singularity \ci{foot14}, 
as we have expected in Sec. \ref{no_Planck}. 
Furthermore, as we will check in the next subsection, 
Eqs. \Ref{sta_metric} and \Ref{eva_metric} correctly contain the effect of 
the four-dimensional Weyl anomaly.
In this sense, they can be regarded as the self-consistent solutions of 
\begin{equation}\lb{Einstein}
G_{\mu\nu}=8\pi G \bra T_{\mu\nu} \ket~.
\end{equation}

We then investigate the behavior of the energy-momentum tensor inside the black hole. 
They can be evaluated from \Ref{sta_metric} for $r\gg l_p$ as \ci{foot16}
\begin{align}\lb{p}
 -\bra T^t{}_t \ket=\f{1}{8\pi G}\f{1}{r^2},&~~~\bra T^r{}_r \ket=\f{1}{8\pi G}\f{1-f}{1+f}\f{1}{r^2}~, \nn \\
 \bra T^\th{}_\th \ket =&\f{1}{8\pi G} \f{1}{2(1+f)^2\s}~.
\end{align}
The energy density $-\bra T^t{}_t \ket=\f{1}{8\pi G}\f{1}{r^2}$ is positive definite everywhere \ci{foot15} 
and consistent with \Ref{rho_a}. 
It gives the mass of the black hole correctly through \Ref{M_formula}:
\begin{equation}\lb{M}
M=4\pi \int^{R(a)}_{\sim \sqrt{\s}} \rd r' r'^2 (-\bra T^t{}_t \ket)\approx \f{a}{2G}~.
\end{equation}
Furthermore, the strong angular pressure $\bra T^\th{}_\th \ket$ appears 
as in \Ref{p_3d}. 
Actually, it breaks the dominant energy condition \ci{Poisson}, 
$\bra T^\th{}_\th \ket\gg -\bra T^t{}_t \ket$, 
and leads to the drastic anisotropy, $\bra T^\th{}_\th \ket\gg \bra T^r{}_r \ket$, 
as we have discussed in Sec. \ref{EMT_fire}. 
Thus, the interior is not a fluid. 

Finally, we check that the energy-momentum flow $P^\mu=\bra T^{\mu \bk}\ket$ 
through the ingoing lightlike spherical surface $S$ 
is equal to the strength $\s$ of the Hawking radiation. 
Actually, the total energy flux measured by the local time is given by
\begin{equation}\lb{J_P}
J \equiv 4\pi r^2 P^\mu u_\mu=4\pi r^2 \f{1}{B}(-\bra T^t{}_t \ket)=\f{\s}{Gr^2}~,
\end{equation}
where $\bm u =e^{-\f{A}{2}}\p_t$, and in the last equation we have used \Ref{B} and \Ref{p}. 
This is consistent with \Ref{da}.

\subsection{Case of conformal matter}\lb{2_5}
In this subsection we consider conformal matter, and show that the 
metric \Ref{sta_metric} is indeed the self-consistent solution of 
$G_{\mu\nu}=8\pi G \bra T_{\mu\nu} \ket$ \ci{KY}. 
We start with the Weyl anomaly \ci{Duff1,BD}
\begin{equation}\lb{anomaly}
G^\mu{}_\mu=8\pi G \bra T^\mu{}_\mu \ket=\g {\cal F}- \a {\cal G} +\f{2}{3} \b \Box  R~, 
\end{equation}
where ${\cal F}\equiv C_{\mu\nu\a\b}C^{\mu\nu\a\b}$ and ${\cal G}\equiv R_{\mu\nu\a\b}R^{\mu\nu\a\b}-4R_{\mu\nu}R^{\mu\nu}+R^2$. 
We have introduced the notations
\begin{equation}\lb{gamma}
\g\equiv 8\pi G \hbar c,~~~\a\equiv 8\pi G \hbar a,~~~\b\equiv 8\pi G \hbar b~,
\end{equation}
where $c,~a,~b$ are the coefficients in the Weyl anomaly. 
This equation together with the assumption \Ref{T2} determines $A(r)$ and $B(r)$ as follows. 

Here, we assume that for $r\gg l_p$, $A(r)$ and $B(r)$ are large quantities of the same order 
as expected from \Ref{B} and \Ref{A}:
\begin{equation}\lb{AB}
A(r)\sim B(r)\gg 1~.
\end{equation}
In order to examine what terms dominate in \Ref{anomaly} for $r\gg l_p$, 
we replace $A$, $B$, and $r$ with $\mu A$, $\mu B$, and $\sqrt{\mu} r$, respectively, 
and pick up the terms with the highest powers of $\mu$. 
Then, we have \ci{foot17}
\begin{align}\lb{mu_eq}
\f{A'^2}{2B}+\cdots&=\g\l(\f{A'^4}{12 B^2}+\cdots  \r)-\a \l(-\mu^{-1}\f{2A'^2}{r^2B}+\cdots \r)\nn\\
&~~~~+\f{2}{3}\b\l(\mu^{-1}\l[\f{A'^3B'}{4B^3}-\f{A'^2A''}{2B^2} \r]+\cdots\r)~.
\end{align}
Therefore, under the assumption $\mu \gg 1$, Eq. \Ref{mu_eq} becomes $\f{A'^2}{2B}=\g\f{A'^4}{12 B^2}$, that is, 
\begin{equation}\lb{B2}
B=\f{\g}{6}A'^2 ~.
\end{equation}
By combining this equation with \Ref{dA}, which is the consequence of \Ref{T2}, 
$\f{2}{1+f}=\f{r\p_r A}{B}~$,
we obtain 
\begin{equation}\lb{AB2}
A'=\f{3(1+f)r}{\g},~~~B=\f{3(1+f)^2r^2}{2\g}~.
\end{equation}

It is natural to expect that the dimensionless function $f(r)$  
is a constant for the conformal fields
\begin{equation}\lb{f_cft}
f(r)={\rm const.}
\end{equation}
If we assume it, we obtain the following interior metric:
\begin{align}\lb{cft}
\rd s ^2 &= -\f{2\g}{3(1+f)^2r^2} e^{-\f{3(1+f)}{2\g}[R(a)^2-r^2]} \rd t^2 \nn\\
&~~~+ \f{3(1+f)^2r^2}{2\g} \rd r^2 + r^2 \rd \Omega^2,~{\rm for}~r\leq R(a).
\end{align}
This contains only two parameters.
One is the $c$-coefficient, which is determined by the matter content.
The other is the constant $f$, which may depend on the detail of the dynamics and
the initial state. 
Thus we have seen that 
\Ref{cft} is the self-consistent solution of $G_{\mu\nu}=8\pi G \bra T_{\mu\nu} \ket$ \ci{foot18}. 

Now, we can understand the origin of the Hawking radiation.
In fact by comparing the second equation in \Ref{AB2} with \Ref{B}, we find 
\begin{equation}\lb{sigma}
\s=\f{\g}{3(1+f)^2}~,
\end{equation}
and the condition for the curvature to be small \Ref{large_N} becomes $\g \gg l_p^2$, that is, 
\begin{equation}\lb{c}
c\gg1~.
\end{equation}
Thus, we have seen that the Hawking radiation is produced by the four-dimensional Weyl anomaly \ci{BD,anomaly,anomaly2}. 
It is interesting that the strength of the Hawking radiation is proportional 
to the $c$-coefficient. 
The positive Hawking radiation ensures the positivity of the $c$-coefficient \ci{foot19}. 
Furthermore, if we compare \Ref{sigma} with \Ref{g2} in the outermost region, 
it is natural to conjecture
\begin{equation}\lb{s0_cft}
\s_0 =\f{\g}{3},~~~1+g=(1+f)^2~.
\end{equation}

We can also understand the existence of the strong angular pressure.
Actually, from \Ref{p}, we have 
\begin{equation}\lb{p_origin}
\bra T^\th{}_\th \ket=\f{1}{2}\bra T^\mu{}_\mu \ket~,
\end{equation} 
which indicates that the large pressure arises as a consequence of the Weyl anomaly.

\subsection{Mechanisms of energy extraction}\lb{3_2}

\subsubsection{Time evolution of the energy of the collapsing matter}
We examine how the collapsing matter loses energy in the 
metric \Ref{sta_metric}. 
In Appendix \ref{Der_e_evo}, we show that the
local energy $\e_{local}(\tau)$ decreases exponentially as a function of
the local time $\tau$:
\begin{equation}\lb{e_evo}
\e_{local}(\tau)=\e_{local}(0)e^{- \f{\tau}{\sqrt{2\s}(1+f)}}~.
\end{equation}
If we consider the outermost region,
the decay time in terms of the proper time 
\begin{equation}\lb{tau_decay}
\Delta \tau_{decay}(r)=\sqrt{2\s(r)}(1+f(r))
\end{equation}
can be converted to the Schwarzschild time
by using $\rd \tau (R(a))\approx \f{\sqrt{2\s(a)}}{a} \rd t$ as
\begin{equation}\lb{t_decay}
\Delta t_{decay}=(1+f(a))a~.
\end{equation}
This is consistent with \Ref{e_decay} except for an extra factor
$(1+f)~$.
The decay time is increased by this factor
because the emission rate is reduced by the scattering.

\subsubsection{Energy exchange and the Weyl anomaly}
Here we analyze the local energy conservation $\N_\a\bra T^\a{}_\b \ket =0$ 
in the presence of the Weyl anomaly, 
and try to get a microscopic picture of the energy decrease \Ref{e_evo}. 
First we express the energy flux from each part by the trace of the
energy-momentum tensor. 
Using the same argument as \Ref{J_shell},
we can show that the outgoing flow of the local energy at $r'$ is given by
\begin{equation}\lb{J_local}
J_{local}(r')=\f{1}{2G}~.
\end{equation}
Then, the local energy flow created by the thin region $r-\Delta r\leq r'\leq  r$ is given by
\begin{align}\lb{DJ_local}
\Delta J_{local}(r) &=J_{local}(r)-J_{local}(r-\Delta r)\sqrt{\f{-g_{tt}(r-\Delta r)}{-g_{tt}(r)}}^2 \nn\\
 &\approx e^{-A(r)}\p_r [e^{A(r)}J_{local}(r)] \Delta r\nn\\
  &=\f{r}{2G(1+f(r))\s(r)}\Delta r~.
\end{align} 
This result is also obtained from the energy conservation and 
the Weyl anomaly.
In fact, by rewriting $\N_\a\bra T^\a{}_\b \ket =0$ for the metric \Ref{sta_metric}, 
we obtain 
\begin{equation}\lb{DJ_local2}
\f{\Delta J_{local}}{\Delta r_{local}}=4\pi \sqrt{2\s}(1+f)\bra T^\mu{}_\mu \ket~,
\end{equation}
where $\Delta r_{local}=\sqrt{g_{rr}}\Delta r$. 
This indicates that 
the outgoing energy is produced at each point by the four-dimensional Weyl anomaly 
and increases as it goes outward. 

Similarly, we can consider the ingoing energy flow. 
From the time reversal symmetry, 
we see that the ingoing energy flow decreases as it goes inward. 
This can happen if
the ingoing energy of the matter is reduced by 
the negative energy that is created from the vacuum inside the black hole. 
This situation is similar to that in the two-dimensional models \ci{BD, 2d_anomaly}. 
In our case, however, the positive energy brought by the collapsing matter is 
greater than the negative energy 
so that the total energy density is positive everywhere as \Ref{p}. 
Thus, the anomaly describes the net effect of the conversion of the ingoing 
energy to the outgoing one. 
In order for this mechanism to work,
there must be proper interactions between the collapsing matter and
the negative energy states.
To describe the precise process of evaporation such as
the information recovery and the baryon number nonconservation,
it should be important to understand such interactions. 


\subsection{Entropy}
\begin{figure}[h]
 \begin{center}
 \includegraphics*[scale=0.18]{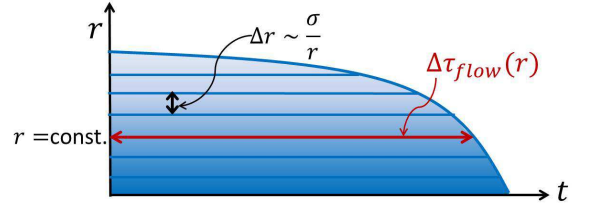}
 \caption{The evaporation of the adiabatically formed black hole and the entropy}
 \label{fig:ent_eva}
 \end{center}
 \end{figure}

Because the adiabatically formed black hole has the maximum entropy,
the area law should be obtained 
if we sum up the entropy of the interior.
More concretely, we put such a black hole in the vacuum, and examine the entropy 
flow during the evaporation.
Because the interior is frozen, 
we slice it to shells with the typical width $\frac{\sigma}{r}$, 
which has been discussed in Sec. \ref{no_Planck}. 
See Fig. \ref{fig:ent_eva}. 
Then, the total entropy should be given by 
\begin{equation}\lb{S1}
S=\int \rd l \int \rd \tau \dot s(r)~,
\end{equation}
where $\dot s (r)$ is the entropy emitted 
per unit proper width $\rd l=\sqrt{g_{rr}}\rd r$ per unit local time $\rd \tau=\sqrt{-g_{tt}}\rd t$. 

We first evaluate the local time $\Delta \tau_{flow}(r)$ 
that it takes for a shell with radius $r$ to appear at the surface,  
\begin{align}\lb{tau_flow}
\Delta \tau_{flow} (r) &=\f{\sqrt{2\s(r)}}{r} \int_0^{t_C} \rd t e^{-\int^{R(a(t))}_{r} \rd r' \f{r'}{2(1+f(r'))\s(r')}} \nn\\
 &\approx \f{\sqrt{2\s(r)}}{r} \int_0^{t_C} \rd t e^{-  \f{a(t)^2-r^2}{4(1+f(r))\s(r)}} \nn\\
 &= \f{\sqrt{2\s(r)}}{r} \int_{r}^{a(0)} \rd a \f{a^2}{2\s (a)} e^{-  \f{a^2-r^2}{4(1+f(r))\s(r)}}\nn\\
 &\approx \sqrt{2\s(r)}(1+f(r))~,
\end{align}
where $t_C$ is the time at which $a(t)$ becomes $r$, $a(t_C)=r$, and \Ref{da} has been used. 
Note that this agrees with \Ref{tau_decay} because time flows only in the
outermost region.

Next, we consider the entropy flux $J_s$ from each shell. 
Because the black hole has been formed adiabatically, we can use
the relation $\Delta S=\f{\Delta Q}{T}$ and have  
$J_s=\f{J_{local}}{T_{local}}$. 
Here, $J_{local}$ is the local energy flux given by \Ref{J_local}, 
and $T_{local}$ is the local temperature at $r$. 
We assume that in the adiabatic process of the black hole formation, 
the local temperature at a point is frozen to the value at the moment 
when the point is covered by the surface. 
Therefore, using $g_{tt}(r)=-1+\f{a'}{r}$, we have
\begin{equation}\lb{T_loc}
T_{local}(a')=\f{1}{\sqrt{-g_{tt}(r=R(a'))}}\f{\hbar}{4\pi a'}= \f{\hbar}{4\pi \sqrt{2\s(a')}}~,
\end{equation}
from which we obtain 
\begin{equation}\lb{J_s}
J_s = \f{J_{local}}{T_{local}}= \f{2\pi \sqrt{2\s(r)}}{l_p^2}~. 
\end{equation}

Then, we can calculate $\dot s$ similarly to \Ref{DJ_local}: 
\begin{align}\lb{dot_s}
\dot s (r) &= \f{1}{\sqrt{g_{rr}}\Delta r } 
\l[J_s(r)-J_s(r-\Delta r) \sqrt{\f{-g_{tt}(r-\Delta r)}{-g_{tt}(r)}}  \r]\nn\\
  &\approx  \f{1}{\sqrt{g_{rr}}} e^{-A(r)}\p_r [e^{A(r)}J_s(r)] \nn\\
 &\approx \f{2\pi}{l_p^2 (1+f(r))} ~.
\end{align}

Now, using \Ref{tau_flow}, \Ref{dot_s}, and $\Delta l (r)=\f{r}{\sqrt{2\s(r)}} \Delta r$, 
we can evaluate \Ref{S1} as  
\begin{align}\lb{S2}
 &~~S_{BH}\nn\\
 =&\int_0^{R(a)} \rd r \f{r}{\sqrt{2\s(r)}} \times \sqrt{2\s(r)}(1+f(r)) \times \f{2\pi}{l_p^2 (1+f(r))}\nn\\
 =&\int_0^{R(a)} \rd r \f{2\pi r}{l_p^2}\approx \f{\pi a^2}{l_p^2}~,
\end{align}
which agrees with the area law, $S_{BH}=\f{A}{4l_p^2}$. 
We can also derive the area law for the stationary black hole in the heat bath . See Appendix \ref{entropy}.

\section{Adiabatically formed charged black hole}\lb{4}
We consider a  Reissner-Nordstrom black hole which is adiabatically formed in the 
heat bath. 
In the following sections we set $\hbar = G=1$. 

\subsection{Test particle near the evaporating charged black hole}
As in the case of the Schwarzschild black hole, 
we start with examining the motion of a test particle near the evaporating Reissner-Nordstrom black hole with mass $M(t)$ and electric charge $Q(t)$. 
We represent the outside spacetime by 
\begin{align}\lb{RN_metric}
 \rd s^2 &=-\f{r^2-2M(t)r+Q(t)^2}{r^2}\rd t^2 \nn\\
 &~~~~+ \f{r^2}{r^2-2M(t)r+Q(t)^2}\rd r^2 + r^2 \rd \Omega^2~,\nn\\
 &=- \f{(r-r_+(t))(r-r_-(t))}{r^2}\rd t^2 \nn\\
 &~~~~+ \f{r^2}{(r-r_+(t))(r-r_-(t))}\rd r^2+ r^2 \rd \Omega^2~, 
\end{align}
where 
\begin{equation}\lb{r+}
r_{\pm }\equiv M \pm \sqrt{M^2-Q^2}~.
\end{equation}
Motivated by the Stephan-Boltzmann law for the entropy flux \ci{foot20}, 
we parametrize the time evolution of $r_+(t)$ as 
\begin{equation}\lb{dr+}
\f{\rd r_+}{\rd t}\equiv  - \f{2\eta_{RN}}{r_+^2},~~~\eta_{RN}\equiv\bar \eta_{RN}\l(\f{r_+-r_-}{r_+}\r)^3~.
\end{equation}
We assume that $\bar \eta_{RN}= \bar \eta_{RN}(M,Q)$ 
is a quantity of order 1 and proportional to the number of fields. 
We also assume that it varies slowly as a function of $M$ and $Q$ and remains finite in the limit $Q\rightarrow M$. 

If a test particle comes sufficiently close to $r_+(t)$, 
its motion is governed by 
\begin{equation}\lb{eom_RN}
\f{\rd r(t)}{\rd t}=-2\kappa_{RN}(r(t)-r_+(t)),~~~\kappa_{RN}\equiv \f{r_+-r_-}{2r_+^2}
\end{equation}
no matter what mass, angular momentum, or electric charge it has. 
Here, $r(t)$ represents its radial position, and $\kappa_{RN}$ is the surface gravity. 
Using a similar argument to \Ref{R0}, 
we obtain 
\begin{equation}\lb{sol_RN}
r(t)\approx r_+ +\f{2\eta_{RN}}{r_+-r_-}+Ce^{-2\kappa_{RN}t}~,
\end{equation}
where we have used \Ref{dr+}. 
Thus, in the time scale of ${\cal O}(r_+)$, any particle reaches
\begin{equation}\lb{R_RN}
R_{RN}(M,Q)\equiv r_+ + \f{2\bar\eta_{RN}}{r_+}\l(\f{r_+-r_-}{r_+}\r)^2~, 
\end{equation}
where \Ref{dr+} has been used. 
The proper distance from the outer horizon is estimated as 
\begin{equation}\lb{dl_RN}
\Delta l =\sqrt{g_{rr}(R_{RN})}\f{2\eta_{RN}}{r_+-r_-}\approx \sqrt{2\bar \eta_{RN}\f{r_+-r_-}{r_+}}~,
\end{equation}
which is larger than the Planck length if we have many fields as in \Ref{largeN_0}.

\subsection{Interior structure and the thermodynamic integrability}
In this subsection, we consider the interior structure of the adiabatically formed charged
black hole.
First, following the same argument as in Sec. \ref{surface2}, from \Ref{R_RN},  
we find that the object has the surface at $r=R_{RN}(M,Q)$ and there is no trapped region inside it. 
Next, we consider the process in which the black hole grows adiabatically 
and discuss how the charge moves in the black hole.
Suppose radiation from the heat bath increases the mass and charge of 
the black hole from $M', Q'$ to $M'+\Delta M, Q'+\Delta Q$.
In this process, 
the electric charge on the original surface $R_{RN}(M',Q')$ should
go outward due to the repulsive force 
and eventually moves to the new surface 
$R_{RN}(M'+\Delta M,Q'+\Delta Q)$.

\begin{figure}[h]
 \includegraphics*[scale=0.16]{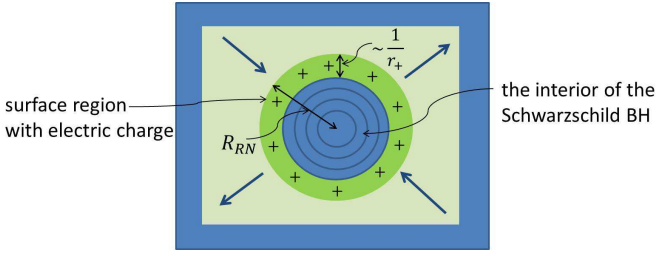}
 \caption{A picture of the Reissner-Nordstrom black hole consistent with 
 the condition for thermodynamical integrability. 
Electric charge exists only in the outermost region with the width $\sim \f{1}{r_+}$.
 } 
 \label{fig:RN}
 \end{figure}

Therefore, it is natural to expect that all the charge is distributed in the
outermost region, where time flows without large redshift.
Then, the interior region is staffed with neutral radiation, and
in accordance with the adiabatic formation, 
it should have the same structure as the interior of the adiabatically 
formed Schwarzschild black hole \Ref{sta_metric}. 
Then there is no inner horizon. 
See Fig. \ref{fig:RN}. 

This picture is natural from the thermodynamic point of view. 
We consider the thermodynamic parameter space $(M,Q)$.
We can move from one point to another along any path 
by controlling the temperature and electric potential of the heat bath.  
If the object really obeys thermodynamics, the state is completely determined
by $(M,Q)$, and should not depend on the path along which  $(M,Q)$ is
reached.
On the other hand, 
the interior of the object is frozen except for the outermost region, as we have seen in section \ref{general}. 
Therefore if the electric charge did not move outward as mentioned above, 
it would be distributed depending on the path, and the thermodynamic
integrability would be violated.

Interestingly, there is a fact that supports this picture. 
Let us consider the energy of the object of Fig. \ref{fig:RN}. 
The inside energy density 
is given by that of the Schwarzschild black hole \Ref{p}, and we obtain 
$\rho_S=\f{1}{8\pi  r^2}$. 
The outside one comes from the electric field of a spherical capacitor of 
radius $r_+$, 
$\rho_{EM}=\f{Q^2}{8\pi r_+^4}$ \ci{Poisson}. 
Thus, using the general formula \Ref{M_formula}, we have 
\begin{align}\lb{M_RN}
 M &\approx 4\pi \int^{R_{RN}}_0 \rd r r^2 \rho_S +4\pi \int^\infty_{R_{RN}} \rd r r^2 \rho_{EM}\nn \\
 &= \f{R_{RN}}{2}+\f{Q^2}{2R_{RN}} \approx \f{r_+}{2}+\f{Q^2}{2r_+}~.
\end{align}
This agrees with \Ref{r+}, which gives a nontrivial
check for the consistency of the above picture.

\subsection{Simple model} 
We now write down the interior metric for such an object. 
Because the charge distribution in the outermost region would be
nontrivial, we assume that the charge is distributed on the surface.
Then, the object consists of 
the interior structure of the Schwarzschild black hole and the charged thin shell. 

Because the inside and outside metrics are the same as that of the Schwarzschild \Ref{sta_metric} and Reissner-Nordstrom
black holes \Ref{RN_metric}, respectively, we have
\begin{widetext}
\begin{equation}\lb{ansatz_RN}
\rd s^2=\begin{cases}
  -\f{2\s(r)}{r^2} e^{- \int^{R_{RN}}_r \rd r' \f{r'}{(1+f(r'))\s(r')}} \rd \tau^2 
+ \f{r^2}{2\s(r)} \rd r^2  + r^2 \rd \Omega^2,~~~{\rm for}~~r\leq R_{RN}~,\\
 - \f{(r-r_+)(r-r_-)}{r^2}\rd t^2 + \f{r^2}{(r-r_+)(r-r_-)}\rd r^2 + r^2 \rd \Omega^2,~~{\rm for}~~r\geq R_{RN}~.
\end{cases}
\end{equation}
\end{widetext}
We will find the relation between two time coordinates $\tau $ and $t$ below.
Here, the surface is considered as a timelike hypersurface located as $r=R_{RN}=$const. 
as in Fig. \ref{fig:surface_sta}. 
Then, the induced metric needs to be connected smoothly \ci{Poisson,Israel}, 
and the condition of the $tt$-component is given by 
\begin{equation}
\f{2 \s(R_{RN})}{R_{RN}^2}\rd \tau^2=\f{(R_{RN}-r_+)(R_{RN}-r_-)}{R_{RN}^2}\rd t^2~, \nn
\end{equation}
which is rewritten by using \Ref{sol_RN} as
\begin{equation}\lb{time_RN}
\rd \tau^2 = \f{\eta_{RN}(M,Q)}{\s(R_{RN})}\rd t^2~.
\end{equation}
Thus, we obtain the metric for the adiabatically formed charged black hole:
\begin{widetext}
\begin{equation}\lb{RN_ful}
\rd s^2=\begin{cases}
  -\f{2\s(r)}{r^2} e^{- \int^{R_{RN}}_r \rd r' \f{r'}{(1+f(r'))\s(r')}} \f{\eta_{RN}(M,Q)}{\s(R_{RN})}\rd t^2 
+ \f{r^2}{2\s(r)} \rd r^2  + r^2 \rd \Omega^2,~~~{\rm for}~~r\leq R_{RN}~,\\
 - \f{r^2-2Mr+Q^2}{r^2}\rd t^2 + \f{r^2}{r^2-2Mr+Q^2}\rd r^2 + r^2 \rd \Omega^2,~~{\rm for}~~r\geq R_{RN}~.
\end{cases}
\end{equation}
\end{widetext}

We can evaluate the energy-momentum tensor on the surface
using Israel's junction condition \ci{Poisson,Israel}. 
In particular, the surface energy density $\e_{2d}$ is given by 
\begin{equation}\lb{e_RN}
\e_{2d} = \f{1}{4\pi}\f{\sqrt{2\s(R_{RN})}}{R_{RN}^2}\l(1-\sqrt{\f{\eta_{RN}(M,Q)}{\s(R_{RN})}}\r)~.
\end{equation}
In order for this to be non-negative, we need to have
\begin{equation}\lb{eta_s}
\eta_{RN}(M,Q)\leq \s(R_{RN})~,
\end{equation}
which means $\rd \tau\leq\rd t$.
This is consistent with the picture that the electric charge lives on the surface. 
Then, it is natural that the $rr$-component of \Ref{RN_ful} jumps at $r=R_{RN}$.

\section{Adiabatically formed slowly rotating black hole}\lb{5}
We discuss a slowly rotating Kerr black hole which is adiabatically formed in the 
heat bath. 
We denote the ratio between ADM energy $M$ and angular momentum $J$ by 
$j\equiv \f{J}{M}$.
\subsection{Test particle near the evaporating Kerr black hole}\lb{SofK}
We begin with analyzing the motion of a test particle near the evaporating 
Kerr black hole. 
A remarkable point is that, although we no longer have the spherical symmetry,
the notion of surface is valid because of the Carter constant.

We assume that the metric of the evaporating Kerr black hole with $M(t)$ and $j(t)$ is given by 
\begin{align}\lb{Kerr}
\rd s^2 &= -\f{\Sigma \Delta}{P} \rd t^2 + \f{P \sin ^2 \th }{\Sigma}(\rd \phi-\o \rd t)^2 + \f{\Sigma}{\Delta} \rd r^2 + \Sigma \rd \th^2~, \nn\\
\Sigma &\equiv r^2 + j(t)^2 \cos^2 \th,~~\Delta \equiv r^2 - 2M(t)r + j(t)^2~,\nn\\
P&\equiv (r^2+j(t)^2)^2-\Delta j(t)^2 \sin^2 \th~.
\end{align}
The angular velocity at $r$ measured by an observer with zero angular momentum is  
\begin{equation}\lb{omega}
\o\equiv \f{2Mj}{P}r~\xrightarrow{r\rightarrow r_+} \f{j}{r_+^2+j^2}\equiv \o_H~,
\end{equation}
where $r_+$ is the location of the outer horizon, which is the solution of $\Delta =0$:
\begin{equation}\lb{Kerr_h}
r_{\pm }\equiv M\pm \sqrt{M^2-j^2}~.
\end{equation}
If a test particle comes close to $r_+(t)$, its motion is described by (see Appendix \ref{Der_K_eom})
\begin{align}\lb{eom_K1}
\f{\rd r(t)}{\rd t}&=-2\kappa_{K}(r(t)-r_+(t))+ {\cal O}(r_+^{-3}),~\kappa_K \equiv \f{r_+-r_-}{2r_+^2}~, \\
\lb{eom_K2}
\f{\rd \th(t)}{\rd t}&={\cal O}(r_+^{-3})~, \\
\lb{eom_K3}
\f{\rd \phi(t)}{\rd t}&=\o_H + {\cal O}(r_+^{-3})  ~, 
\end{align}
no matter what mass and angular momentum it has. 
The test particle approaches $r_+$ in the time scale of ${\cal O}(r_+)$.
It rotates with the black hole while it hardly moves in the $\th$-direction. 
Using a similar analysis to Sec. \ref{surface}, we obtain
\begin{equation}\lb{sol_K}
r(t)\approx r_+ - \f{1}{2\kappa_{K}}\f{\rd r_+}{\rd t}+Ce^{-2\kappa_{K}t}~.
\end{equation}
Therefore, any particle approaches
\begin{equation}\lb{R_K}
R_{K}(M,j)\equiv r_+ - \f{1}{2\kappa_{K}} \f{\rd r_+}{\rd t}~,
\end{equation}
and we conclude that there is a clear surface at $r=R_{K}(M,j)$.

\subsection{Slowly rotating limit}
We now investigate the interior of the adiabatically formed Kerr black hole. 
However, it is difficult to analyze the general Kerr black hole because 
the discussions in Sec. \ref{general} depend on the spherical symmetry. 
In the following parts, as a first trial, 
we consider the slowly rotating limit 
in which we ignore terms with ${\cal O}(j^2)$. 
Then the Kerr metric \Ref{Kerr} reduces to 
\begin{align}\lb{Kerr2}
\rd s^2 &= - \f{r-2M(t)}{r}\rd t^2 + \f{r}{r-2M(t)}\rd r^2 \nn\\
&+ r^2 \rd \th ^2 + r^2 \sin^2 \th[\rd \phi - \o(t,r) \rd t]^2 + {\cal O}(j^2) 
\end{align}
with 
\begin{equation}\lb{omega2}
\o=\f{2Mj}{r^3} + {\cal O}(j^2),~r_+=2M+{\cal O}(j^2),~\kappa_K=\f{1}{4M}+ {\cal O}(j^2)~.
\end{equation}
Now the location of the horizon and the surface gravity are the same as the Schwarzschild black hole. 
If we introduce another angular coordinate by
$\rd \psi \equiv \rd \phi -\o(r,t) \rd t$, 
Eq. \Ref{Kerr2} becomes
\begin{align}\lb{Kerr3}
\rd s^2 &= - \f{r-2M(t)}{r}\rd t^2 + \f{r}{r-2M(t)}\rd r^2 \\
 &+ r^2 (\rd \th ^2 +\sin^2 \th \rd \psi^2) + {\cal O}(j^2) ~. \nonumber
\end{align}
Note that a trajectory with $\psi=$const. rotates with the angular velocity $\f{\rd \phi}{\rd t}=\o(r,t)$ with respect to the static coordinate at infinity. 
Therefore, we can regard this object simply as the Schwarzschild black hole rotating with 
the angular velocity 
\begin{equation}\lb{omega_H2}
\o_H=\f{j}{r_+^2}+{\cal O}(j^2)~.
\end{equation}

Then, we can understand that
the object emits energy in the same way as the Schwarzschild black hole, which is consistent with the
emission of angular momentum. 
Hence, from \Ref{da}, \Ref{R_K}, and \Ref{omega2}, we have
\begin{equation}\lb{R_K2}
R_K(M)=r_+ + \f{2\s(r_+)}{r_+}~+ \cO(j^2)
\end{equation}
for $r_+=2M+\cO(j^2)$, which is the same as \Ref{R}. 

\subsection{Simple model}
We now investigate the interior metric of such an object.
Because \Ref{Kerr3} is spherically symmetric and the surface is
given by \Ref{R_K2},
we can follow the discussion of Sec. \ref{surface2}. 
As in the case of the charged black hole, 
the thermodynamic integrability in $(M,j)$ space requires that 
the angular momentum is distributed only in the outermost region 
with width $\sim \f{1}{r_+}$ and that
the interior is the same as the Schwarzschild black hole. 
Here, we assume that due to the centrifugal force the
angular momentum moves outward at each step of the adiabatic growth. 
Thus, we obtain a picture similar to Fig. \ref{fig:RN} in which electric charge is 
replaced by angular momentum, and the whole system is rotating.

In order to write down the metric, 
we model the object by composing the interior of the Schwarzschild black hole 
and a thin layer with the angular velocity \Ref{omega_H2}.
Then, from \Ref{sta_metric} and \Ref{Kerr3}, we obtain the metric 
\begin{widetext}
\begin{equation}\lb{Kerr_in}
\rd s^2=\begin{cases}
-\f{2\s(r)}{r^2} e^{- \int^{R_K(M)}_r \rd r' \f{r'}{(1+f(r'))\s(r')}} \rd t^2 + \f{r^2}{2\s(r)} \rd r^2 + r^2 [\rd \th ^2 + \sin^2 \th \rd \psi_H^2],~~{\rm for}~~r\leq R_K(M)~,\\
 - \f{r-2M}{r}\rd t^2 + \f{r}{r-2M}\rd r^2 + r^2 [\rd \th ^2 + \sin^2 \th (\rd \phi - \o(r)\rd t)^2] + {\cal O}(j^2),~~{\rm for}~~r\geq R_K(M)~,
\end{cases}
\end{equation}
\end{widetext}
where 
$\rd \psi_H \equiv \rd \phi -\o_H \rd t$. 

As a consistency check, we evaluate the energy-momentum tensor on the surface $r=R_K$ by using Israel's junction condition \ci{Poisson, Israel}. 
We obtain the surface energy density $\e_{2d}$, surface pressure $p_{2d}$, and surface angular momentum density ${\cal J}_{2d}$: 
\begin{align}
\e_{2d}=0,~&~p_{2d}=-\f{1}{16\pi}\f{1-f(R_K)}{1+f(R_K)}\f{1}{\sqrt{2\s(R_K)}}~, \nn\\
{\cal J}_{2d}&=\f{3j\sin^2 \th}{16\pi\sqrt{2\s(R_K)}}~.
\end{align}
These reproduce the angular momentum $J$ through the generalized Komar formula \ci{Cruz}, 
while there is no additional contribution to the ADM energy $M$ from the surface.

\section{Conclusions and discussions}\lb{6}
In this paper we have considered time evolution of the black hole 
with the backreaction from the Hawking radiation taken into account. 
We have found that a collapsing matter becomes an object that looks like
a black hole when it is seen from the outside.
However, instead of the horizon, it has a clear boundary, which we call
the surface.
The inside of it is filled with matter and radiation,
while the outside is almost empty.
The surface is located slightly outside the horizon of the vacuum solution.
For example, in the spherically symmetric case, it is located at $r=a+\f{2\s(a)}{a}$, 
where $a$ is the Schwarzschild radius.
Because the structure inside the surface is totally different from the 
vacuum solution, the object has neither a trapped region nor singularity.
In general, the inside structure depends on the initial distribution
of the matter, because time evolution is almost frozen inside the surface 
due to the large redshift.
However, if we see the object from the outside, it looks the same
as the conventional picture of the black hole:
it emits the Hawking radiation 
and evaporates in the time scale $\sim \f{a^3}{\s}$. 

In this sense, the black hole in the real world
is not the vacuum region with the closed trapped surface, 
but a kind of highly dense star. 
Therefore, the problem of the time evolution of the information is similar 
to that of the wave function of a many-particle system in condensed matter physics.
Here, it is important to consider interactions. 
Actually, we can estimate the time scale in which the information 
comes back in the evaporation process by considering 
the interaction between the matter and the Hawking radiation. 

However, there remain problems to be clarified. 
First, our description is based on the spherical symmetry. 
It is important to extend it to the general case 
such as the construction of the interior metric of the rotating black hole 
and the investigation of the nonspherical symmetric instability of \Ref{cft}. 

Another important problem is to understand 
how baryon number changes through interactions in the evaporation process. 
We have seen that it should occur in the outermost region, 
where the energy scale of the particles is close to the Planck scale but still controllable 
by field theory if we have many species of fields. 
It is interesting to see whether some effects of quantum gravity 
or string theory are involved in the mechanism or not.

We also have not identified the microscopic mechanism of 
the strong angular pressure, which supports the object. 
Although we have found that it is necessary to satisfy the Weyl anomaly,
we have not understood how it occurs.
This problem is also related to how the incoming energy of the 
collapsing matter 
is converted to the outgoing energy of the Hawking radiation.

It is also interesting to construct more explicit relations 
among the phenomenological functions, $\s_0(a),~g(a),~f(a)$, by considering a simple example. 
Then, we can see explicitly how the intensity $\s(a)$ 
depends on the initial data of the matter. 

Finally, it is attractive to investigate the process in which the charged and 
rotating black hole 
is formed from the matter with a general distribution and 
check the validity of the thermodynamic integrability more explicitly.
\section*{Acknowledgment}
The authors would like to thank members in ICTS and S. Sasa for valuable discussions. 
This work has been refined by the comments and discussions in 
Bangalore Area String Meeting at ICTS, IAGRG 2015 at Raman Research Institute,  
KEK Theory Workshop 2015 at KEK, Strings 2015 at ICTS, 
and a seminar at Indian Institute of Astrophysics. 
Y.Y. thanks Department of Physics in Kyoto University and Yukawa Institute for hospitality. 

\appendix 

\section{Black hole entropy and Bekenstein's argument}\lb{Bek_rev}
Based on Bekenstein's gedankenexperiment, 
we discuss black hole entropy. 
First we estimate information to be lost in the formation process. 
Then we discuss the area law, and argue that 
generically the entropy production occurs in the evaporation process. 
Finally we review an operation for massive particles. 

\subsection{Entropy in the formation process}
Bekenstein introduced the notion of black hole entropy 
as the logarithm of the number of all the possible ways $\Omega$ to construct the black hole \ci{Bekenstein}. 
We review and generalize his idea. 
Suppose we construct a black hole with radius $a$ from matter. 
First, we focus on the stage where the radius is $a_i$ 
and throw a particle with energy $\e_i$ to the black hole. 
Here, in order for the particle to enter into the black hole, 
Eq. \Ref{e_B0} needs to be satisfied. 
Let us estimate the number of all the possible ways $\Omega_i$ for this process. 
Note that the wavelength of the particle is $\lam_i\sim \f{\hbar}{\e_i}$, 
which plays a role of the spatial resolution for this process. 
Hence, the number of the ways for the particle to enter into the black hole is 
given by 
\begin{equation}\lb{O_i}
\Omega_i \sim \f{a_i}{\lam_i} \times n_i \sim \f{a_i\e_in_i}{\hbar}~. 
\end{equation}
Here, $n_i$ is the number of species of particles with $\e_i$, which is assumed to be $\cO(1)$. 
Equation \Ref{O_i} corresponds to the phase volume for a particle with energy $\e_i$ and species $n_i$
in a one-dimensional system with size $\sim a_i$, 
because the black hole is spherically symmetric. 

Applying this estimation to each stage of the construction, 
we obtain
\begin{equation}\lb{O}
\Omega = \prod_{i=1}^N \Omega_i\sim \prod_{i=1}^N \f{a_i\e_in_i}{\hbar}~.
\end{equation}
In order to evaluate the total number of steps $N$, we consider the case where 
$a_i\sim a$, $\e_i\sim \e$, and $n_i\sim n$ for any $i$. 
Then, we have 
\begin{equation}\lb{N}
N\sim \f{a}{G\e}
\end{equation}
because the size and energy of the black hole are related as $a=2GM$. 
Thus, Eq. \Ref{O} becomes 
\begin{equation}\lb{OO}
\Omega \sim  \l(\f{a \e}{\hbar}n\r) ^N ~, 
\end{equation}
which leads to 
\begin{equation}\lb{S_gen}
S_{BH}^{formation}=\log \Omega \sim \f{a}{G\e}\log \l(\f{a\e }{\hbar}\r)~.
\end{equation}
Here, the contribution of $\log n$ is neglected 
because it is smaller than $\log \l(\f{a\e }{\hbar}\r)$ due to \Ref{e_B0}. 
By construction, this entropy measures the 
amount of information that is lost in the formation process. 
This is because the information of the matter which has entered into the horizon is lost 
for an outside observer. 

We consider here the saturating case in \Ref{e_B0}, that is, 
the case of \Ref{e_B}. 
This is the most slow formation that corresponds to the adiabatic 
process in the sense of thermodynamics. 
Then, Eq. \Ref{S_gen} becomes 
\begin{equation}\lb{S_adia}
S_{BH}^{formation}\sim \f{a^2}{l_p^2}\log \cO(1)\sim \f{a^2}{l_p^2}~, 
\end{equation}
which agrees with the conventional area law \ci{Hawking,G-H}
\begin{equation}\lb{Sformula}
S_{BH}^{formula}=\f{A}{4l_p^2}~,
\end{equation}
except for the coefficient. 

Here, we discuss the relation with the new picture of black holes. 
As is discussed in Sec. \ref{general}, in the new picture, 
the interior of the black hole is almost
frozen due to the large redshift, 
and the information of the initial distribution is kept inside for long time.
Eventually it will come out in the process of evaporation,
but in practice it seems to be lost.
In this sense, the above discussion applies also
to the new picture.

\subsection{Entropy production in the evaporation process}
We discuss the relation between the general result \Ref{S_gen} 
and the usual formula \Ref{Sformula} and show that generically the entropy 
production occurs in the evaporation process. 
Originally, Eq. \Ref{Sformula} was obtained by integrating
the Hawking radiation from the evaporating black hole \ci{Hawking}. 
In this sense, it counts the number of microstates in the matter emitted from the black hole. 
On the other hand, 
Eq. \Ref{S_gen} measures the number of all possible ways to construct the black hole. 
Note that \Ref{Sformula} is much larger than \Ref{S_gen} unless $\e\sim \f{\hbar}{a}$. 
Therefore, in the whole process from formation to evaporation, 
the entropy increases by 
\begin{equation}\lb{}
\Delta S= S_{BH}^{formula}-S_{BH}^{formation}\sim \f{a^2}{l_p^2}- \f{a}{G\e}\log \l(\f{a\e }{\hbar}\r)~, 
\end{equation}
which is consistent with the generalized second law \ci{Bekenstein}.
Especially, if we consider the adiabatic formation \Ref{S_adia}, 
$\Delta S\sim 0$, which corresponds to 
the black hole in equilibrium with the heat bath of the Hawking temperature \ci{G-H}.

\subsection{``Reversible" process of massive particles}
From the above discussion it seems that 
a massive particle with $\e>\f{\hbar}{a}$ cannot be given to the black hole adiabatically, 
since $\Delta S>0$ for $\e>\f{\hbar }{a}$. 
However, we can make a procedure to add a massive particle to the black hole 
``reversibly." 

Suppose we give a massive particle with rest mass $m\gg \f{\hbar}{a}$, say, a proton with $m\sim 1GeV$. 
If we throw it far from the black hole, the ADM energy increases by $\e=m$. 
Now, we follow Bekenstein's gedankenexperiment \ci{Bekenstein}.
Imagine that we slowly lower the massive particle from a point at $r\gg a$ to the black hole 
by using a strong string. 
If the string is cut off at a point $r$, the ADM energy given to the black hole is evaluated as 
\begin{equation}\lb{binding_e}
\e(r)=m\sqrt{1-\f{a}{r}}~,
\end{equation}
which is smaller than $\e(r=\infty)=m$. 
This is because in this process the system composed of the black hole and the particle 
works positively to an external agent by the gravitational binding energy. 
Here, we consider a question: 
In order to give the maximum entropy and the minimum energy to the black hole  by this process, 
how far do we have to lower the particle? 
In quantum mechanics a massive particle exists within the Compton wavelength $\lambda _C=\f{\hbar}{m}$. 
Therefore, if the particle is located at a distance of its Compton wavelength from the horizon, 
we cannot distinguish whether the particle already has been swallowed into the black hole or not. 
This situation leads to loss of 1 bit information for an outside observer. 
Indeed, we can see this explicitly as follows. For $r=a+\Delta r$, 
Eq. \Ref{binding_e} becomes 
\begin{equation}
\e(r=a+\Delta r)\approx m \sqrt{\f{\Delta r}{a}}=m\f{l}{2a}~,
\end{equation}
where \Ref{l} has been used. 
Therefore, by taking $l=\lambda_C=\f{\hbar}{m}$ we have 
\begin{equation}\lb{e_C}
\e(r=a+\Delta r)=\f{\hbar}{2a}~,
\end{equation}
which agrees with the minimum energy \Ref{e_B}. 
Thus, if we repeat this process to make a black hole, 
no entropy production occurs in the evaporation process: $\Delta S\sim 0$. 
More rigorously, however, the baryon number should be changed in the 
evaporation process as is discussed in Sec. \ref{baryon}, 
and the entropy production should occur.

Finally, we show that when such a massive particle released at $r=a+\Delta r (l=\lam_C)$ 
comes close to the surface at $r=R(a)=a+\f{2\s}{a}$, it becomes ultrarelativistic, and
can be regarded as an ingoing lightlike particle.
In fact its local energy is estimated as  
\begin{align}
\e_{local}&=m\f{\sqrt{-g_{tt}(r=a + \Delta r|_{\lambda _C})}}{\sqrt{-g_{tt}(r=R)}}\nn\\
 &\approx m\sqrt{\f{\f{\Delta r|_{\lambda _C}}{a}}{\f{2\s}{a^2}}} 
 =m\f{a}{\sqrt{2\s}}\f{\lambda_C}{2a}=\f{\hbar}{2\sqrt{2\s}}
\end{align}
where \Ref{l} has been used. 
This is the same order as \Ref{e1_loc} and much larger than the rest mass $m$. 

\section{A numerical solution of $r(t)$} 
\begin{figure}[h]
\begin{center}
\includegraphics*[scale=0.16]{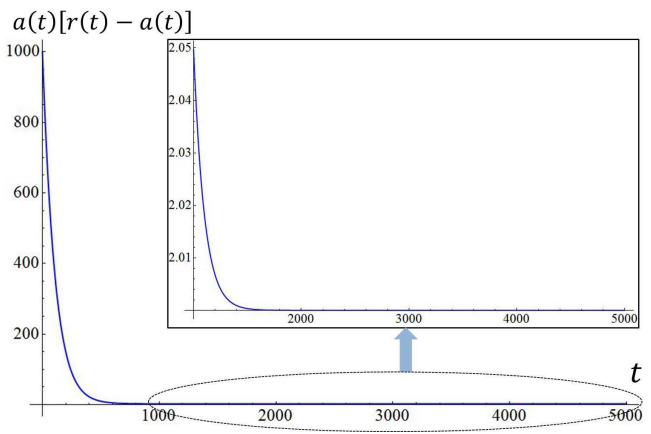}
\caption{The numerical result of $a(t)[r(t)-a(t)]$ for $a(0)=100,~r(0)=110,~\sigma=k=1$. 
It approaches $2k=2$.}
\label{fig:r_plot}
\end{center}
\end{figure}

\lb{R_numerical}
We give a numerical demonstration of \Ref{R0}. 
We consider a simple case where $\sigma(a)=k=$const. in \Ref{da}, 
the solution of which is given by $a(t)=[a(0)^3-6kt]^{1/3}$. 
Using this, we solve \Ref{r_t} numerically for $a(0)=100,~r(0)=110$, and $k=1$ 
and obtain Fig. \ref{fig:r_plot}. 
This shows that 
$r(t)$ approaches $a(t)+\f{2k}{a(t)}$ as in \Ref{R0}. 

\section{A multishell model}\lb{many_model}
We introduce a simple model \ci{KMY} as a concrete example for the new picture in Sec. \ref{story}. 

\subsection{Single-shell model}
\begin{figure}[h]
\begin{center}
\includegraphics*[scale=0.18]{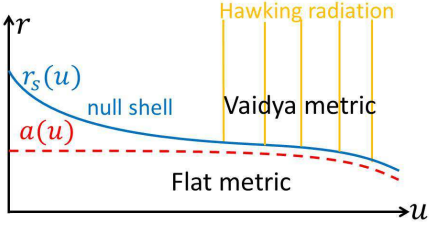}
\caption{An evaporating null shell. 
In this coordinate $(u,r)$, an outgoing null ray is depicted by a line with $u=$const.. 
}
\label{fig:1-shell}
\end{center}
\end{figure}

In order to construct the self-consistent solution of $G_{\mu\nu}=8\pi G\bra T_{\mu\nu}\ket$, 
we start with a spherically symmetric collapsing null shell. 
The inside region is flat because of the spherical symmetry.  
We assume that the outside region is described by the outgoing Vaidya metric \ci{Vaidya}:
\begin{equation}
\rd s^2=\begin{cases}
- \rd U^2 -2\rd U \rd r + r^2\rd \Omega^2~~{\rm for}~~r\leq r_s\\
- \f{r-a(u)}{r}\rd u^2 -2\rd u \rd r + r^2\rd \Omega^2~~{\rm for}~~r\geq r_s
\end{cases}
\end{equation}
Here $r_s$ is the radius of the shell, $M(u)=\f{a(u)}{2G}$ is the Bondi mass, and 
$U$ and $u$ are the outgoing null coordinates in the Eddington-Finkelstein coordinates for the inside and outside, respectively. 
See Fig. \ref{fig:1-shell}.

In Ref.\ci{KMY}, we have shown that 
the self-consistent solution is obtained by solving 
\begin{equation}\lb{uU}
\f{r_s(u)-a(u)}{r_s(u)}\rd u=-2\rd r_s = \rd U
\end{equation}
and 
\begin{equation}\lb{a_eom_A}
\f{\rd a}{\rd u}= - \f{l_p^2}{8\pi}\{u,U \}~,
\end{equation}
provided that the radiation is massless, and only the s-wave is considered in the eikonal approximation. 
Here, $\{u,U \}$ is the Schwarzian derivative, 
which is given by 
$\{u,U \}\equiv \f{\ddot U(u)^2}{\dot U(u)^2}-\f{2\dddot U(u)}{3\dot U(u)}$. 
Equation \Ref{uU} indicates that 
the motion of the shell is lightlike both in the inside and the outside regions. 
Equation \Ref{a_eom_A} is equivalent to the Einstein equation $G_{\mu\nu}=8\pi G\bra T_{\mu\nu}\ket$ 
because the only nonzero components of $G_{\mu\nu}$ and $\bra T_{\mu\nu}\ket$ 
are $G_{uu}=-\f{\dot a(u)}{r^2}$ and $\bra T_{uu}\ket=\f{1}{4\pi r^2}\f{\hbar}{16\pi} \{ u,U\}$.

\subsection{Generalization to multishells}
\begin{figure}[h]
\begin{center}
\includegraphics*[scale=0.21]{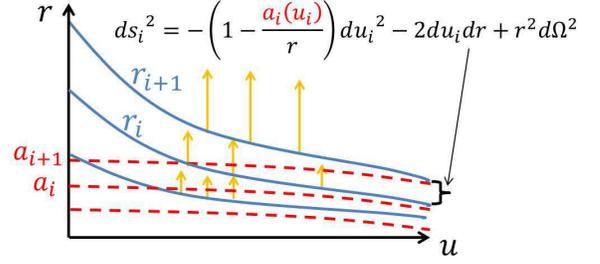}
\caption{A continuous distribution modeled by many shells. }
\label{fig:many_shell}
\end{center}
\end{figure}

It is easy to generalize the above model to the case of multishells. 
We consider $n$ shells of which the radii are denoted by $r_i$ $(i=1,...n)$.  
See Fig. \ref{fig:many_shell}. 
The metric of the region between $r_i$ and $r_{i+1}$ is described by 
\begin{equation}\lb{Vaidya_i}
\rd s^2_i = - \f{r-a_i(u_i)}{r}\rd u_i^2 -2\rd u_i \rd r + r^2\rd \Omega^2~,
\end{equation}
where $u_i$ is the local time and $\f{a_i(u_i)}{2G}$ is the mass in the $i$th shell. 
The junction condition of each time coordinate $u_i$ is given as \Ref{uU} by 
\begin{equation}\lb{uU_i}
\f{r_i-a_i}{r_i}\rd u_i=-2\rd r_i = \f{r_i-a_{i-1}}{r_i} \rd u_{i-1}~~{\rm for}~~i=1,...n~, 
\end{equation}
where we regard $a_0=0$ and $u_0=U$ for the flat space. 
This is equivalent to 
\begin{equation}\lb{ri_eom}
\f{\rd r_i}{\rd u_i}=- \f{r_i-a_i}{2r_i}~,
\end{equation}
and 
\begin{equation}\lb{ui_eom}
\f{\rd u_i}{\rd u_{i-1}}=\f{r_i-a_{i-1}}{r_i-a_i}=1+\f{a_i-a_{i-1}}{r_i-a_i}~.
\end{equation}
The Einstein equation \Ref{a_eom_A} holds for each shell, 
\begin{equation}\lb{a_eom_i}
\f{\rd a_i}{\rd u_i}= - \f{N l_p^2}{8\pi}\{u_i,U \}~,
\end{equation}
where the degrees of freedom of the fields $N$ have been introduced. 

The coupled equations \Ref{uU_i} and \Ref{a_eom_i} can be solved by the following ansatz: 
\begin{equation}\lb{ansatz1}
\f{\rd a_i}{\rd u_i}= - \f{C}{a_i^2},
\end{equation}
\begin{equation}\lb{ansatz2}
r_i=a_i-2a_i\f{\rd a_i}{\rd u_i}=a_i + \f{2C}{a_i}~.
\end{equation}
Equation \Ref{ansatz1} means that each shell behaves like the conventional evaporating black hole, 
and \Ref{ansatz2} indicates that 
each shell has reached the asymptotic radius as in \Ref{R0}. 

First, Eq. \Ref{ansatz2} solves \Ref{ri_eom} as in \Ref{R0}. 
Next, we solve \Ref{ui_eom} with the ansatz.
We define $\eta_i$ by 
\begin{equation}
\eta_i\equiv \log \f{\rd U}{\rd u_i}~.
\end{equation}
Then, we have 
\begin{align}
\eta_i-\eta_{i-1} &=\log \f{ \f{\rd U}{\rd u_i} }{ \f{\rd U}{\rd u_{i-1}}} =- \log \f{\rd u_i}{\rd u_{i-1}}\nn\\
 &=-\log \l( 1+\f{a_i-a_{i-1}}{r_i-a_i} \r)  \nn\\
 &\approx -\f{a_i-a_{i-1}}{r_i-a_i} \approx -\f{a_i-a_{i-1}}{\f{2C}{a_i}} \nn\\
 &\approx - \f{1}{4C} \l(a_i^2-a_{i-1}^2\r)~.
\end{align}
Here, at the second line, we use \Ref{ui_eom}; 
at the third line, we assume $\f{a_i-a_{i-1}}{\f{2C}{a_i}}\ll 1$, 
which is satisfied for the case of continuous distribution; 
and at the last line, we approximate $2a_i\approx a_i+a_{i-1}$. 
With the boundary conditions $\eta_0=a_0=0$, we obtain
\begin{equation}\lb{eta_result}
\eta_i =-\f{1}{4C}a_i^2.
\end{equation}

Finally, we check \Ref{a_eom_i}. 
Because the Schwarzian derivative $\{u_i,U\}$ is written as 
\begin{equation}\lb{Sch_formula}
\{u_i,U\}=\f{1}{3}\l(\f{\rd \eta_i}{\rd u_i}\r)^2-\f{2}{3}\f{\rd^2 \eta_i}{\rd u_i^2}~,
\end{equation}
we obtain 
\begin{equation}\lb{uU_result}
\{u_i,U\}\approx \f{1}{12}\f{1}{a_i^2}~.
\end{equation}
Here, we have used \Ref{ansatz1} and \Ref{eta_result} to obtain 
\begin{equation}
\f{\rd \eta_i}{\rd u_i} =-\f{1}{2C}a_i \f{\rd a_i}{\rd u_i}=\f{1}{2a_i}~.
\end{equation}
Therefore, Eq. \Ref{a_eom_i} is satisfied if
\begin{equation}\lb{C}
C=\f{Nl_p^2}{96\pi}.
\end{equation}

In particular, the outermost shell ($n$th shell) satisfies 
\begin{equation}
\f{\rd a}{\rd u}=-\f{C}{a^2},
\end{equation}
which indicates that 
the entire system behaves like the conventional evaporating black hole 
when it is observed from the outside.


\section{Hawking radiation in the new picture}
\begin{figure}[h]
\includegraphics*[scale=0.16]{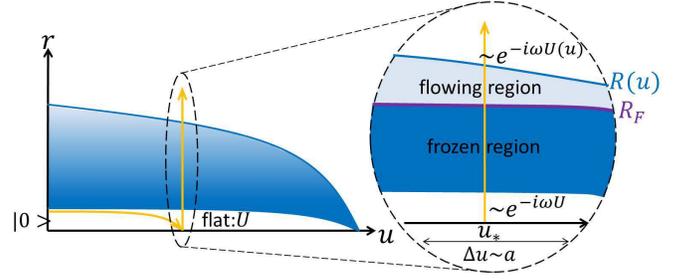}
\caption{A continuously distributed matter that collapses and evaporates 
and a trajectory of a field from the flat spacetime to the matter. 
Here, an outgoing null line is depicted as $u=$const.}
\label{fig:creation}
\end{figure}

\lb{Der_T_H} 
We show that the metric of the new picture of black holes
indeed creates radiation from the vacuum which obey a Planck-like distribution
with the Hawking temperature \Ref{T_H} \ci{foot21}. 
We start with the general form of the spherically symmetric metric
in the Eddington-Finkelstein-like coordinates: 
\begin{equation}\lb{hq_metric}
\rd s^2 =-q(u,r)\l(\f{h(u,r)}{r}\rd u + 2\rd r \r)\rd u + r^2 \rd \Omega ^2~.
\end{equation}
See Fig. \ref{fig:creation}. 
We assume that the metric near the surface, $R_F<r\leq R$, is given 
by \Ref{eva_metric}. 
Rigorously speaking, the values of $f$ and $\sigma$ there for the evaporating
black hole may be different from those of the adiabatically formed one. 
Here we simply assume that they are the same. 
On the other hand, the deeper region, $0 \lesssim r \leq R_F$, need not be the same 
as  \Ref{eva_metric}, but the metric there is frozen in time.
We further assume that there remains a small flat spacetime around the origin: 
\begin{equation}\lb{flat_A} 
\rd s ^2 = - \rd U^2 -2\rd U \rd r + r^2 \rd \Omega^2~. 
\end{equation}

We consider the s-wave of a massless real scaler field $\phi(x)$ and 
solve the Heisenberg equation using the eikonal approximation \ci{Hawking}. 
In this approximation, the reflection of the radiation is not considered, 
and $f(r)$ becomes zero. 
The Hawking radiation is created by the time evolution of the field along a ray
that starts from the flat space before the collapse,
goes through the center, and passes the collapsing matter (see Fig. \ref{fig:creation}). 
We focus here on a time interval around $u_*$, 
$I(u_*)=[u_*-ka(u_*),u_*+ka(u_*)]$, where $k\sim 1$. 
By putting $\phi=r^{-1}e^{i\f{S(u,r)}{\hbar}}$, 
the Klein-Gordon equation
$\N^2 \phi = 0$ in the metric \Ref{hq_metric} becomes
\begin{equation}
\l[\f{1}{q}\p_u - \f{h}{2rq}\p_r\r]S~\p_rS=0~,
\end{equation}
in the leading approximation of the $\hbar$ expansion.
Here, the ingoing and outgoing radial null vectors are given by 
$\bm k = \f{1}{q}\p_u - \f{h}{2rq}\p_r$ and $\bm l = \p_r$, respectively. 
Therefore, we obtain the outgoing eikonal solution 
\begin{equation}\lb{S_out}
\phi_{out}=\f{e^{i\f{S_{out}(u)}{\hbar}}}{r}~. 
\end{equation}

We assume that the field was in the vacuum state before the collapse.
Therefore, in the Heisenberg picture, the state should satisfy 
\begin{equation}\lb{vac_A}
a_{\o}|0\ket =0~,
\end{equation}
where $a_{\o}$ is the coefficient of $e^{-i\o U}$ for $\o>0$
in the outgoing component of $\phi~$:
\begin{equation}
\phi= \int_0^\infty \f{\rd \o}{\sqrt{2\pi}}\f{1}{\sqrt{2\o}}
\l( \f{e^{-i\o U}}{\sqrt{4\pi}r}a_\o +  \f{e^{i\o U}}{\sqrt{4\pi}r}a_\o^\dagger \r)~. 
\end{equation}
Because of \Ref{S_out}, the field which has evolved is given by 
\begin{equation}\lb{phi_after}
\phi= \int_0^\infty \f{\rd \o}{\sqrt{2\pi}}\f{1}{\sqrt{2\o}}
\l( \f{e^{-i\o U(u)}}{\sqrt{4\pi}r}a_\o +  \f{e^{i\o U(u)}}{\sqrt{4\pi}r}a_\o^\dagger \r)~. 
\end{equation}

On the other hand, this can also be expressed in terms of 
the modes in the future infinity: 
\begin{equation}\lb{phi_b}
\phi= \int_0^\infty \f{\rd \o}{\sqrt{2\pi}}\f{1}{\sqrt{2\o}}
\l( \f{e^{-i\o u}}{\sqrt{4\pi}r}b_\o +  \f{e^{i\o u}}{\sqrt{4\pi}r}b_\o^\dagger \r)~. 
\end{equation}
Generally, two operators $a_\o$ and $b_{\o'}$ are related as 
\begin{equation}\lb{b_a}
b_\o=\int^\infty_{-\infty}\rd \o' A_{\o\o'} a_{\o'}= 
\int^\infty_{0}\rd \o'( A_{\o\o'} a_{\o'} + A_{\o,-\o'} a_{\o'}^\dagger )~. 
\end{equation}
From \Ref{phi_after}, \Ref{phi_b}, and \Ref{b_a}, 
we obtain 
\begin{equation}\lb{coe_A}
A_{\o,-\o';u_*}= \f{1}{2\pi}\sqrt{\f{\o}{\o'}}\int_{-\infty}^{\infty}\rd u 
e^{i\o u}e^{i\o' U(u)} ~.
\end{equation}
Here, we have extended the integration region from $I(u_*)$ to $[-\infty,\infty]$, 
which does not affect the following calculation \ci{Hawking, KMY, Barcelo} \ci{foot22}. 
From \Ref{b_a}, we obtain the expectation value of the number of the particles:
\begin{equation}\lb{bb}
\bra 0| b_\o^\dagger b_\o | 0\ket_{u_*} = \int_0^\infty \rd \o'|A_{\o,-\o';u_*}|^2~. 
\end{equation}

Then, we estimate $U(u)$, which is the relation between 
the outside time $u$ and the time $U$ around the origin.
In the metric \Ref{hq_metric}, the ingoing radial geodesic $r(u)$ is given by 
$\f{1}{q}\f{\rd r}{\rd u}=- \f{h}{2rq}$, 
and $q(u,r(u))\rd u$ corresponds to the local time at $r=r(u)$. 
Because $u$ is the outside time and $q(u, r=R)=1$, we have
\begin{align}\lb{U_u0}
\f{\rd U(u)}{\rd u}&=q(u, r\sim 0)=\f{q(u,r\sim 0)}{q(u, r=R_F)} q(u, r=R_F)\nn\\
 &=C e^{-\int^{R(u)}_{R_F}\rd r\f{r}{2\s(r)}}\approx C e^{-\f{R(u)^2-R_F^2}{4\s(R(u))}}~.
\end{align}
Here, in the interval $I(u_*)$, the deeper region between $0\lesssim r \leq R_F$ is 
frozen in time so that $\f{q(u,r\sim 0)}{q(u, r=R_F)}$ is a constant, $C$. 
On the other hand, 
the region near the surface, $R_F<r\leq R(u)$, is described by the interior metric of \Ref{eva_metric} with $f=0$. 
Next, we expand $R(u)$ around $u_*$ as 
\begin{align}
R(u)&\approx R(u_*) + \f{\rd R}{\rd u}(u_*)(u-u_*) \nn\\
 &= R(u_*)- \f{\s(R(u_*))}{R(u_*)^2}(u-u_*)~,
\end{align}
where we have used the geodesic equation in \Ref{eva_metric}, $\f{\rd R}{\rd u}=-\f{\s(R)}{R^2}$. 
Putting $R(u)^2 \approx R(u_*)^2- \f{2\s(R(u_*))}{R(u_*)}(u-u_*)$ into \Ref{U_u0}, 
we have $\f{\rd U(u)}{\rd u}\approx C' e^{\f{u}{2R(u_*)}}$. 
Thus, we obtain 
\begin{equation}\lb{U_u}
U(u;u_*)= D +2R(u_*)C' e^{\f{u}{2R(u_*)}}~, 
\end{equation}
where $D$ and $C'$ are constants. 

Now, we can evaluate \Ref{coe_A}. 
Using \Ref{U_u} and continuing analytically \ci{Hawking}, we have
\begin{equation}
A_{\o,-\o';u_*} = \f{1}{2\pi}\sqrt{\f{\o}{\o'}}2R(u_*) e^{-\pi \o R(u_*)}\Gamma (2i \o R(u_*)) ~,
\end{equation}
where an irrelevant overall phase factor has been dropped. 
Then, employing the formula $|\Gamma(ix)|^2=\f{\pi}{x \sinh (\pi x)}$ for $x \in \mathbb{R}$ 
and considering wave packets in \Ref{bb} \ci{Hawking,Barcelo}, 
we obtain 
\begin{equation}
\bra 0| b_\o^\dagger b_\o | 0\ket_{u_*} = \f{1}{e^{\f{\hbar \o}{T_H(u_*)}}-1}~, 
\end{equation}
where 
\begin{equation}
T_H(u)=\f{\hbar}{4\pi R(u)}\approx \f{\hbar}{4\pi a(u)}~. 
\end{equation}
This agrees with the Hawking temperature \ci{Hawking} 
but changes in time slowly according to \Ref{da}. 
The above analysis has shown that 
the Hawking radiation is created
in any collapsing process 
as long as the region near the surface becomes 
the asymptotic spacetime described by \Ref{eva_metric}.
In particular, the existence of the horizon is not necessary.

\section{Examples of the interior structure}
\begin{figure}[h]
\begin{center}
\includegraphics*[scale=0.125]{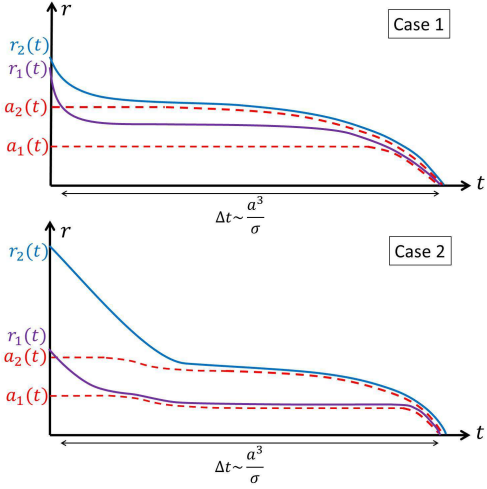}
\caption{Examples of two-shell collapsing process.}
\label{fig:general3}
\end{center}
\end{figure}

\lb{ex_eva}
Simple examples would be helpful to understand the new picture. 
We consider two macroscopic shells with radii $r_1$ and $r_2$, where $r_1<r_2$.
We assume initially they have the same energy $M$. 
Then, the Schwarzschild radius of the total system and the inner shell are
$a_2=4GM$ and $a_1= 2GM$, respectively.
Although in the real system the energy distributes continuously,
for simplicity we consider two thin shells.
See Fig. \ref{fig:general3}.

We consider the following two cases.
In one case the initial values of the radii are very close, for example,
$r_1= 3a_1$ and $r_2= 3a_1 + \Delta r$, where $\Delta r \ll a_1$ 
(see the upper panel of Fig. \ref{fig:general3}). 
Then, shell 2 reaches $r=R(a_2)$ earlier than shell 1 reaches $r=R(a_1)$
because the two shells have started from 
almost the same positions but $R(a_2)$ is significantly larger than $R(a_1)$. 
When shell 2 reaches $r=R(a_2)$, time inside shell 2 is frozen, 
and $r_1$ stays almost constant.
At the same time, shell 2 
starts emitting the Hawking radiation. 
Thus, only shell 2 evaporates until 
its energy is exhausted. 
After that, shell 1 begins to 
approach to $R(a_1)$ and eventually evaporates. 

In the other case, shell 1 reaches $r=R(a_1)$ earlier than 
shell 2 reaches $r=R(a_2)$.
Such a case occurs if the initial value of $r_1$ is close to $R(a_1)$ but
$r_2$ is not so close to $R(a_2)$ (see the lower panel of Fig. \ref{fig:general3}). 
Then, shell 1 reaches $R(a_1)$ before it is frozen by shell 2,
and some part of it evaporates 
until shell 2 reaches $R(a_2)$. 
Then, shell 1 stops radiating, 
and only shell 2 emits the energy. 
After this, the same thing occurs as in the first case.

\section{Surface energy density and pressure on the evaporating shell}\lb{Israel_con}
We derive \Ref{EMT_shell} and \Ref{p_2d_shell} 
from the Barrabes-Israel junction condition \ci{Israel_null} 
following the formalism in Ref.\ci{Poisson}. 
We consider an evaporating shell with initial energy $\e(0)\sim \f{\hbar}{a}$ 
which approaches the evaporating core with $a'(t')$ as in Fig. \ref{fig:general}.  
The metric is given by \Ref{Sch2}, 
and we denote the position of the shell by $r_s$. 
We take the ingoing and outgoing null vectors as $\bk = f^{-1}\p_T-\p_r$ and $\bl = \f{1}{2}\p_T+\f{f}{2}\p_r$, 
respectively, so that $\bk\cdot\bl=-1$. 
Here, $T=t$ and $f=1-\f{a}{r}$ for $r>r_s$, 
while $T=t'$ and $f=1-\f{a'}{r}$ for $r<r_s$. 
Note that the null vector $\bk$ satisfies the geodesic equation 
$\N_{\bk}k^\mu=\kappa k^\mu$ with 
\begin{equation}\lb{k_kappa}
\kappa = -f^{-2}\p_T f~.
\end{equation}

We assume that the shell moves along $\bk$ 
and denote its locus by ${\cal N}$. 
The transverse curvature of the null hypersurface ${\cal N}$ is given by 
\begin{equation}
C_{ab}=\f{1}{2}(\pounds_{\bl}g_{\mu\nu})e^\mu_a e^\nu_b.
\end{equation}
Here $\pounds_{\bl}$ is the Lie derivative along $\bl$, 
and $\{\bm e_a\}=\{\bm e_{\lam}=\bk, \bm e_{\th}=\p_\th, \bm e_{\varphi}=\p_\varphi \}$ 
is the basis on ${\cal N}$. 
Using another expression $C_{ab}=-l_\nu e^\mu_a\N_\mu e^\nu_b$ (see Ref. \ci{Poisson}), 
we obtain 
\begin{equation}\lb{C_results}
C_{\lam\lam}=\kappa,~C_{AB}=\f{f}{2r}\s_{AB}~, 
\end{equation}
where $\s_{AB}$ is the 2-metric on the shell 
such that $\s_{AB}\rd \th^A \rd \th^B =r^2 \rd \th^2 +r^2\sin^2\th\rd\varphi^2$ 
with $\th^A=\{\th,\varphi\}$. 
Thus, we obtain the formula \Ref{EMT_shell} for the surface energy density $\e_{2d}$ and pressure $p_{2d}$: 
\begin{align}\lb{p_2d_A}
\e_{2d} &=-\f{1}{8\pi G}\s^{AB}[C_{AB}]=\f{\e}{4\pi r_s^2}~, \nn\\
p_{2d} &=  -\f{1}{8\pi G}[C_{\lam\lam}] =-\f{1}{8\pi G}[\kappa] \nn\\
       &=-\f{r_s}{8\pi G (r_s-a)^2}\l(\f{\rd a}{\rd t}-\l(\f{r_s-a}{r_s-a'} \r)^2 \f{\rd a'}{\rd t'} \r)~.
\end{align}
Here, $[A]\equiv A|_{r\to r_s+0}-A|_{r\to r_s-0}$ for a quantity $A$, 
and $\e=\f{a-a'}{2G}$ is the energy of the shell. 

Next, we show \Ref{p_2d_shell}. 
Suppose that the shell has come close to the core as in stage II of Fig. \ref{fig:general}, 
where $r_s=a+\f{2\s(a)}{a}$. 
Using \Ref{largeN_0}, \Ref{tt'2}, $\Delta a \sim \f{l_p^2}{a}$, and $\f{1}{a'^2}\approx\f{1}{a^2}\l(1+\f{2\Delta a}{a}\r)$, 
$p_{2d}$ in \Ref{p_2d_A} becomes 
\begin{align}
&p_{2d}\nn\\
\approx&\f{a}{8\pi G}\f{a^2}{(2\s)^2}\l(\f{2\s(a)}{a^2}-\l(1-2\f{a\Delta a}{2\s(a)}\r)\f{2\s(a)}{a^2}\l(1+\f{2\Delta a}{a}\r) \r)\nn\\
 \approx& \f{a^2\Delta a}{16\pi G \s^2}\nn\\
 \sim& \f{a}{GN^2l_p^2}~.
\end{align}

\section{Exponential decrease in the energy of the collapsing matter}\lb{Der_e_evo}
Here we derive \Ref{e_evo}. 
We consider ingoing matter with radius $r=r(t)$
and width $\Delta r(t)$, 
in the outermost region of the evaporating black hole with radius $a(t)$.
Here we assume the metric is given by \Ref{sta_metric} 
and examine the time evolution of the ADM energy of the shell $\e(t)$, 
which is expressed as
\begin{equation}\lb{e_tau}
\e(t )=4\pi r(t)^2 \rho (r(t)) \Delta r(t)=\f{1}{2G}\Delta r(t)~,
\end{equation}
where $\rho = -\bra T^t{}_t\ket =\f{1}{8\pi G r^2}$ because of \Ref{p}.

We first consider the motion of a test particle in \Ref{sta_metric}. 
If it is ultrarelativistic, it satisfies 
\begin{equation}\lb{eomAB}
\f{\rd r}{\rd t} = - \f{1}{B} e^{\f{A}{2}}~.
\end{equation}
Taking the difference of this equation with respect to $r$, we obtain
\begin{equation}\lb{eom_dr0}
\f{\rd \Delta r}{\rd t}=\l(\f{\p_r B}{B}-\f{1}{2}\p_r A \r)\f{e^{\f{A}{2}}}{B} \Delta r
 \approx - \f{r}{2\s (1+f)}\f{e^{\f{A}{2}}}{B} \Delta r ~,
\end{equation}
where we have used \Ref{A} and \Ref{B}.
This can be rewritten in terms of the local time 
$\rd \tau = \f{1}{\sqrt{B(r)}}e^{\f{A(r)}{2}}\rd t$,
as 
\begin{equation}\lb{eom_dr}
\f{\rd \Delta r}{\rd \tau} = - \f{1}{\sqrt{2\s}(1+f)}\Delta r~,
\end{equation}
and the solution is given by 
\begin{equation}\lb{dr_evo}
\Delta r(\tau)=\Delta r (0)e^{- \f{\tau}{\sqrt{2\s}(1+f)}}~.
\end{equation}
Therefore, the time evolution of 
the local energy $\e_{local}=\f{\e}{\sqrt{-g_{tt}}}$ is given by 
\begin{equation}
\e_{local}(\tau)=\e_{local}(0)e^{- \f{\tau}{\sqrt{2\s}(1+f)}}~.
\end{equation}

\section{Entropy of the stationary black hole}\lb{entropy}
We consider the stationary black hole that is formed adiabatically
in the heat bath and evaluate the total entropy by summing up
the contribution from each piece in the interior.
Here, we approximate 
the interior as one-dimensional massless radiation with local equilibrium. 
Because 
the local energy of particles is ultrarelativistic,
and the ingoing and outgoing energy flux are balanced 
at each $r$,  
it is reasonable to consider such a model.

Then, the Gibbs relation for one-dimensional radiation \ci{Landau_S} 
should be satisfied, 
\begin{equation}\lb{Gibbs}
u+p=T_{local}s,~~~u=p~,
\end{equation}
where $u,~p,~T_{local}$, and $s$ are one-dimensional internal energy density, 
pressure, local temperature, and entropy density, respectively. 
In the following, we evaluate
\begin{equation}\lb{s}
s=2 \f{u}{T_{local}}
\end{equation}
and integrate it over the inside of the black hole by using the metric \Ref{sta_metric}. 

We first evaluate $u$. 
The general mass formula \Ref{M_formula} indicates that $\rho=-\bra T^t{}_t\ket$ 
can be regarded as the four-dimensional energy density in the local inertial frame. 
This is because the integration measure is the same as the ordinary flat space, 
and the diagonal component $T^t{}_t$ is the same as $T^\tau{}_\tau$ 
for the stationary state. 
Then, the one-dimensional energy density, that is, the proper energy per 
proper length, is given by \ci{foot23} 
\begin{equation}\lb{e}
e=4\pi r^2 \rho~.
\end{equation}
Here, by noting that the ingoing and outgoing energy 
flows are balanced in the stationary state, 
the energy flow only in one direction should be considered to count the 
entropy in \Ref{s}. 
Therefore, by using \Ref{p}, 
we can evaluate the one-dimensional internal energy density as 
\begin{equation}\lb{u}
u=\f{1}{2}e=\f{1}{4G}. 
\end{equation}

Now we can evaluate the entropy. 
From \Ref{s}, \Ref{u}, and \Ref{T_loc}, we obtain
\begin{equation}\lb{s2}
s(r)=2\pi\f{\sqrt{2\s(r)}}{l_p^2}~,
\end{equation}
which indicates that $\sqrt{\s}/l_p^2$ bits are stored per unit proper 
length, as discussed in Ref. \ci{foot9}. 
Integrating this from $r\sim 0$ to $r=R(a)=a+\f{2\s(a)}{a}$, we have 
\begin{align}\lb{S_BH}
S_{BH}&=\int_0^{R(a)} \rd r \sqrt{g_{rr}(r)} s(r) \nn\\
 &= \int_0^{R(a)} \rd r \f{r}{\sqrt{2\s(r)}} 2\pi\f{\sqrt{2\s(r)}}{l_p^2} \nonumber \\
 &\approx \f{\pi a^2}{l_p^2}=\f{A}{4l_p^2}~,
\end{align}
which agrees with the Bekenstein-Hawking formula. 
Thus, we have seen that the black hole entropy is stored in the interior
structure.

\section{Surface of the rotating black hole}\lb{Der_K_eom}
In Sec. \ref{SofK} we analyze the motion of a test particle in the 
evaporating Kerr metric,
and show that the notion of surface is valid even for the rotating
black holes.
In this Appendix, we derive the three equations \Ref{eom_K1},
\Ref{eom_K2} and \Ref{eom_K3}, which play a crucial role
for the existence of the surface.

If the black hole is not very close to extremal, 
the relevant time scales are similar to those of the 
spherically symmetric black holes. 
The time scale of the change of the mass $M(t)$ and the 
angular momentum $j(t)$ is about $M^3$, 
while the test particle approaches the horizon in
the time scale of $\sim M$.
Therefore it is enough to analyze the equation of motion of the test particle
for constant $M$ and $j$, and then replace them to the time-dependent ones.

In order to do that, we consider the Hamilton-Jacobi equation 
for a massive particle with mass $m$ \ci{Poisson} in the Kerr metric \Ref{Kerr}: 
\begin{equation}\lb{HJ}
0=g^{\mu\nu}\p_\mu S \p_\nu S + m^2.
\end{equation}
As usual, because of the cylindrical symmetry, we can set
\begin{equation}\lb{HJ_S}
S(t,r,\th,\phi)=-Et + L \phi + W(r,\th)~.
\end{equation}
By using the inverse metric 
\begin{align*}
 &g^{tt}=-\f{P}{\Sigma \Delta},~g^{rr}=\f{\Delta }{\Sigma},~g^{\th \th}=\f{1}{\Sigma}, \\
&g^{\phi \phi}=\f{\Delta -j^2 \sin^2 \th}{\Sigma \Delta \sin^2 \th},~g^{t\phi}=-\f{2Mjr}{\Sigma \Delta }~,
\end{align*}
Eq. \Ref{HJ} becomes 
\begin{align}\lb{HJ1}
 0&=-\f{P}{\Sigma \Delta} E^2 + \f{\Delta }{\Sigma} (\p_r W)^2 + \f{1}{\Sigma} (\p_\th W)^2  \nn\\
 &+ \f{\Delta -j^2 \sin^2 \th}{\Sigma \Delta \sin^2 \th} L^2
+2\f{2Mjr}{\Sigma \Delta} E L +m^2~.
\end{align}
Multiplying this by $\Sigma$, we obtain 
\begin{widetext}
\begin{align}\lb{HJ2}
0 &= \l( -\f{1}{\Delta}[(r^2+j^2)^2E^2+j^2L^2-4MjrLE]+\Delta(\p_r W)^2 +(r^2+j^2)m^2 \r) \nn\\
 &~~~~~~~~~~~~~+ \l(j^2 \sin ^2 \th E^2 + \f{L^2}{\sin ^2 \th} + (\p_\th W)^2 - j^2 \sin ^2 \th m^2 \r)~. 
\end{align}
The first and second terms depend only on $r$ and $\th$, respectively. 
Therefore, we can put the first term as a constant, $-C$, where $C$ is the Carter constant \ci{Poisson}, 
and decompose $W(r,\th)$ as 
\begin{equation}\lb{HJ_W}
W(r,\th) = w(r) + \Theta (\th)~. 
\end{equation}
Then we have the two ordinary differential equations and obtain the solutions 
\begin{equation}\lb{sol_w}
w(r;E,L,C) = \pm \int^{r} \rd r \f{1}{\Delta}\sqrt{(r^2+j^2)^2E^2+j^2L^2-4MjrLE- \Delta[C+(r^2+j^2)m^2] }~, 
\end{equation}
\end{widetext}
\begin{equation}\lb{sol_th}
\Theta(\th;E,L,C) =\pm \int^{\th} \rd \th \sqrt{ C-\f{L^2}{\sin^2\th}-j^2\sin^2\th (E^2-m^2)}~.
\end{equation}
Taking two derivatives of 
\begin{equation}\lb{S_sol}
S=-Et+L\phi +w(r;E,L,C)+\Theta(\th;E,L,C)
\end{equation}
with respect to $E$ and to $t$, we have
\begin{align}\lb{r_th}
0&=-1\pm \f{(r_+^2 + j^2)}{\Delta (r(t))}\f{\rd r(t)}{\rd t} \nn\\
&\pm \f{-E j^2 \sin^2 \th(t)}{\sqrt{C-\f{L^2}{\sin^2\th(t)}-j^2\sin^2\th(t) (E^2-m^2)}}\f{\rd \th(t)}{\rd t}~.
\end{align}

We now focus on the region $r\sim r_+$, where $\Delta \sim 0$.
Using $\Delta=(r(t)-r_+)(r(t)-r_-)\approx (r(t)-r_+)(r_+-r_-)$ and 
$\kappa_K=\f{r_+-r_-}{2(r_+^2 + j^2)}$, from \Ref{r_th}
we obtain
\begin{align}\lb{r1}
&\f{\rd r(t)}{\rd t}=-2\kappa_K (r(t)-r_+)\nn\\
&~~~~\times\l[1\pm \f{E j^2 \sin^2 \th(t)}{\sqrt{C-\f{L^2}{\sin^2\th(t)}-j^2\sin^2\th(t) (E^2-m^2)}}\f{\rd \th(t)}{\rd t} \r],
\end{align}
where we have chosen the sign for the radial ingoing direction. 
Similarly, taking two derivative of \Ref{S_sol} with respect to $C$ and $t$, 
we obtain 
\begin{equation}\lb{th1}
\f{\rd \th(t)}{\rd t}
= \f{\sqrt{C-\f{L^2}{\sin^2\th(t)}-j^2\sin^2\th(t) (E^2-m^2)}}{(r_+^2 + j^2)E -jL} \f{\rd r(t)}{\rd t}~.
\end{equation}
Then substituting this into \Ref{r1}, we have 
\begin{equation}\lb{r2}
\f{\rd r(t)}{\rd t}=-2\kappa_K (r(t)-r_+)\l[1\pm \f{E j^2 \sin ^2 \th(t)}{(r_+^2 + j^2)E -jL} \f{\rd r(t)}{\rd t} \r]~. 
\end{equation}

The second term in the large bracket in $\Ref{r2}$ turns out to be
of order ${\cal O}(r_+^{-2})$, 
and it can be neglected compared with the first term. 
We can show this as follows, 
Since we consider adiabatic processes, the ingoing energy $E$ should be of the
same order as the Hawking temperature, $E\sim T_H \sim \f{1}{r_+}$.
The typical angular momentum of the ingoing particles can be estimated as 
$L\sim$(impact parameter)$\times$(energy)$\sim r_+ E \sim 1$. 
Furthermore, as in the spherically symmetric case, we have
\begin{equation}
r(t)-r_+(t)\sim \f{1}{r_+(t)}~,
\end{equation}
which will be shown self-consistently.
Using these estimates, we can show the above statement.

Thus, we obtain 
\begin{equation}\lb{r3}
\f{\rd r(t)}{\rd t}=-2\kappa_K (r(t)-r_+) + {\cal O}(r_+^{-3})~,
\end{equation}
which is \Ref{eom_K1}. 
From this, Eq. \Ref{th1} can be estimated as 
\begin{equation}
\f{\rd \th(t)}{\rd t}\sim \f{1}{r_+}2\kappa_K (r(t)-r_+) \sim \f{1}{r_+^3}~,
\end{equation}
which gives \Ref{eom_K2}. 

Finally we examine $\f{\rd \phi(t)}{\rd t}$. 
Taking two derivatives of \Ref{S_sol} with respect to $L$ and $t$, 
we have 
\begin{align*}
 0&=\f{\rd \phi(t)}{\rd t} \pm \f{-a}{\Delta} \f{\rd r(t)}{\rd t}\nn\\
&~~~\pm \f{-\f{L}{\sin^2 \th(t)}}{\sqrt{C-\f{L^2}{\sin^2\th(t)}-j^2\sin^2\th(t) (E^2-m^2)}} \f{\rd \th(t)}{\rd t} \\
 &=\f{\rd \phi(t)}{\rd t} \pm \f{-a}{\Delta} \f{\pm \Delta}{r_+^2 + j^2} + {\cal O}(r_+^{-3})~.
\end{align*}
Thus, we have \Ref{eom_K3}:
\begin{equation}
\f{\rd \phi(t)}{\rd t} =\o_H+{\cal O}(r_+^{-3})~.
\end{equation}



\begin{thebibliography}{99}
\bibitem{foot1}
We will examine the definition of $\s(a)$ more precisely in Sec. \ref{2_6}.
\bibitem{foot2}
It gives the energy flow \Ref{da} correctly, $-G^r{}_t=\f{1}{r^2}\f{\rd a}{\rd t}$,  
but gives the zero energy density, $G^t{}_t=0$, 
although the radiation around the black hole should have the energy density of order
$-\bra T^t{}_t\ket\sim \f{N}{\hbar^3}T_H^4\sim \f{N\hbar}{a^4}$. ($N$ is the number of species of fields.) 
However, the effect of the latter is so small that the motion of the test particle is not altered. 
We can show \ci{KMY} that the result is not changed even if we take
the outgoing Vaidya metric \ci{Vaidya} instead of \Ref{Sch} (see also Appendix \ref{many_model}).
In the case of the stationary black holes in the heat bath, we can regard
\Ref{Sch} with $a=$const. as the approximate solution.
\bibitem{foot3}
See Appendix \ref{Der_K_eom} for the precise argument. 
There we also find that $\f{\rd \th(t)}{\rd t}\sim\f{\rd \phi(t)}{\rd t} =\cO\l(\f{l_p^2}{a^3}\r)$, 
which indicates that near $r\sim a$ particles move only in the radial direction.
\bibitem{foot5}
This notion of boundary is different from the stretched horizon 
in the membrane paradigm \ci{membrane} in that we consider the inside of the boundary as well as the outside.
\bibitem{foot5_5}
In Sec. \ref{2_5} we will check that this result is consistent with the four-dimensional Weyl anomaly.
\bibitem{foot6}
In Appendix \ref{ex_eva}, 
we give simple examples of the time evolution of collapsing matter.
\bibitem{foot7}
Several authors have obtained Penrose diagrams similar to Fig. \ref{fig:Penrose_eva} \ci{works}. 
However, they have considered the black hole 
as a compact vacuum region surrounded by a trapped surface, which is different from our approach.
There are similar works in which the collapsing objects are
described from the point of view of an outside observer
before a horizon is formed \ci{works2}.
\bibitem{foot8} 
The detailed mechanism will be discussed in the next section.
\bibitem{foot9}
From this, we can conclude that the entropy per unit proper length is given
by $s\sim \f{\s/l_p^2}{\Delta l} \sim \f{\sqrt{\s}}{l_p^2}$.
In Appendix \ref{entropy}, 
we check that this is integrated to reproduce the entropy area law.
\bibitem{foot12}
In Ref \ci{KMY}, based on this idea, 
a many-shell model is used to construct the self-consistent interior metric, 
which is essentially the same as \Ref{sta_metric}.
\bibitem{foot13}
In this sense, the ordinary evaporation of the black hole in the vacuum is not completely adiabatic
because $a$ changes by $\Delta a \sim \f{\s}{a}$ in the time scale $a$ [as \Ref{e_decay}], which
is shorter than  $a\log \f{a}{l_p}$. 
We can also estimate the entropy production (see Appendix \ref{Bek_rev}).
\bibitem{foot14}
This can also be understood as follows. 
If we focused on $r\sim 0$, the leading behavior would be given by 
$R\sim \f{1}{r^2},~~~\sqrt{R_{\mu\nu}R^{\mu\nu}},~\sqrt{R_{\mu\nu\a\b}R^{\mu\nu\a\b}}\sim \f{\s}{r^4}$, 
which would diverge at $r=0$, but the macroscopically large quantity $a$ 
does not appear in the coefficients. 
More concretely, 
we estimate the size of the world line along $r={\rm const.}\sim l_p$ 
for the evaporating black hole with $a(t)$: 
$\int \sqrt{-\rd s^2} \sim \int \rd t e^{-\int^{a(t)}_{l_p} \rd r' \f{r'}{2(1+f(r'))\s(r')}}   
 \sim \int e^{-\f{a(t)^2}{4(1+f)\s}}\rd t = \int \rd a \f{\rd t}{\rd a} e^{-\f{a^2}{4(1+f)\s}}
 \sim \f{1}{\s}\int_0^{\infty}\rd a a^2 e^{-\f{a^2}{4(1+f)\s}} \sim \sqrt{\s}$. 
From this and \Ref{curve}, we find that
the curvature radius and lifetime of the would-be timelike singularity around the origin 
is small and controllable in the sense that 
it can be removed by introducing a string of the size $\sim \sqrt{\s}$ surrounding the origin. 
This contrasts to the macroscopically large spacelike singularity of the vacuum Schwarzschild 
black hole solution, $\sqrt{R_{\mu\nu\a\b}R^{\mu\nu\a\b}} \sim \f{a}{r^3}$.
\bibitem{foot16}
Note here that the energy-momentum tensor jumps at $r=R(a)$ 
while the Israel junction condition is satisfied \ci{Poisson,Israel}. 
Such a rapid change would become smooth if we consider the time evolution 
of the outermost region more precisely.
\bibitem{foot15}
This is singular at $r=0$. However, as we have discussed in \Ref{large_N}, 
this solution is valid only in the region $\sqrt{\s}\lesssim r\leq R(a)$, 
where $-\bra T^t{}_t \ket\lesssim \f{1}{G\s}$ is not trans-Planckian.
\bibitem{foot17}
We assume that the coefficients of the higher-curvature terms 
in the action are 
renormalized to order 1. 
However, $\a$ and $\g$ are proportional to the degrees of freedom $N$ because 
they are not canceled by counterterms.
\bibitem{foot18}
We can also check that the exterior metrics in \Ref{sta_metric} and \Ref{eva_metric} are well-approximated solutions of the
self-consistent equation including the Weyl anomaly because we have
$8\pi G \bra T^\mu{}_\mu \ket \sim \f{l_p^2 a^2}{r^4} \approx 0$  
and $G^\mu{}_\mu=0$ for $r>a$.
\bibitem{foot19}
This is consistent with a discussion based on the spectral representation \ci{spectral}.
\bibitem{foot20}
It is given by 
$\f{\rd S}{\rd t}\propto-A T^3$ \ci{Landau_S}. 
Using $A=4\pi r_+^2$, $T=\f{\kappa_{RN}}{2\pi}$, and $S=\pi r_+^2$, we have \Ref{dr+}.
\bibitem{foot21}
A similar derivation is done in Ref. \ci{KMY}. 
However, we consider here the general interior metric.
\bibitem{foot22}
Equation \Ref{coe_A} is the same as that of the Klein-Gordon inner product \ci{BD} 
if we use such a Cauchy surface that includes as a part the surface $R(u)$ for $I(u_*)$.
\bibitem{foot23}
We can check this directly from the Hamiltonian
$H = \int dr 4\pi r^2 \sqrt{-g_{tt}g_{rr}}\f{1}{2}[-g^{tt}(\p_t \phi)^2+g^{ii}(\p_i \phi)^2] \equiv \int dr {\cal H}_r$, 
where ${\cal H}_r$ is the Hamiltonian density per unit radial coordinate.
Using
$\rho\equiv -T^t{}_t=\f{1}{2}[-g^{tt}(\p_t \phi)^2+g^{ii}(\p_i \phi)^2]$,
we have 
${\cal H}_r=4\pi r^2 \sqrt{-g_{tt}g_{rr}}\rho$.
Therefore, the one-dimensional energy density $e$ is given by 
$e=\f{d E_{local}}{d l_{local}}=\f{{\cal H}_r}{\sqrt{-g_{tt}g_{rr}}}=4\pi r^2 \rho$, 
because $d E_{local}=\f{{\cal H}_r }{\sqrt{-g_{tt}}}dr$, and $dl_{local}=\sqrt{g_{rr}}dr$.
\end{thebibliography}
\end{document}